# A Modelling Framework for Regression with Collinearity

Takeaki KARIYA [1], Nagoya University of Commerce and Business, Graduate School of Management, Nisshin, Aichi, Japan 470-0193
The corresponding author: thekariya70@gmail.com
Hiroshi KURATA[2]   University of Tokyo,  Graduate School of Arts and Sciences, Meguroku, Tokyo, Japan 153-8902  kurata@waka.c.u-tokyo.ac.jp
Takaki HAYASHI[3]   Keio University, Graduate School of Business Administration, Yokohama, Kanagawa, Japan 223-8526  hayashi.kbs@gmail.com

Abstract

This study addresses a fundamental, yet overlooked, gap between standard theory and empirical modelling practices in the OLS regression model $y = X\beta + u$ with collinearity. In fact, while an estimated model in practice is desired to have stability and efficiency in its "individual OLS estimates",  $y$ itself has no capacity to identify and control the collinearity in $X$ and hence no theory including model selection process (MSP) would fill this gap unless $X$ is controlled in view of sampling theory.  In this paper, first introducing a new concept of "empirically effective modelling" (EEM), we propose our EEM methodology (EEM-M) as an integrated process of two MSPs with data ($y^o, X$) given. The first MSP uses $X$ only, called the XMSP, and pre-selects a class  $D$ of models with individually inefficiency-controlled and collinearity-controlled OLS estimates, where the corresponding two controlling variables are chosen from predictive standard error of each estimate. Next, defining an inefficiency-collinearity risk index for each model, a partial ordering is introduced onto the set of models to compare without using $y^o$, where the better-ness and admissibility of models are discussed.  The second MSP is a commonly used MSP that uses ($y^o, X$), and evaluates total model performance as a whole by such AIC, BIC, etc. to select an optimal model from $D$. Third, to materialize the XMSP, two algorithms are proposed.

# 1 Introduction

Within the traditional ordinary least squares (OLS) framework in the linear regression model with collinearity

(1.1) $\quad y = X\beta + u$ with $E(u) = 0$ and $Var(u) = \sigma^2 I$,

this paper first proposes a new concept of "Empirically Effective Modelling" (EEM) together with the EEM methodology (EEM-M) for finding an empirically effective model from an observation $(y^o, X)$ for model in (1.1). The EEM-M is based on the EEM concept and forms a new integrated model selection process (MSP) consisting of *bundle model* concept, two different MSPs, new model comparison methodology and algorithms for selecting inefficiency-controlled and collinearity-controlled models, where the two MSPs are the XMSP with using $X$ only and $y^o$MSP with using $(y^o, X)$. The whole picture of the EEM-M is symbolically described as an integrated process of $B\mathcal{M}([X]) - \text{XMSP} - y^o\text{MSP}$. Here $B\mathcal{M}([X])$ denotes a bundle model replacing (1.1) and it makes the two MSP concepts consistent with the EEM (see Section 1.2 for details). Also, the XMSP includes a decision-theoretic framework of comparing models in terms of the inefficiency risk and the collinearity risk (see Sections 1.4 and 3).

As notation, let $X = (x_{nk}) = (x_1, x_2, \cdots, x_K)$ with $x_1 = (1, \cdots, 1)' \equiv e$ be an $N \times K (N > K)$ explanatory matrix of rank $K$, and $X$ is assumed to contain all possible variables for MSP and to be used as they are. We call each vector $x_k$ a "variable" with $x_k = \{x_{nk}\}$, and $X$ a "model". And $X$ is interchangeably used as the set consisting of $K$ vectors $\{x_k\}$ as well. Similarly, a sub-set $X^*$ of $X$ is called a sub-model or simply a model formed as a matrix and it is assumed that $X^*$ always includes a constant term, i.e., $x_1 = e$. Throughout this paper the sample size $N$ is fixed.

Our concept of collinearity is here confirmed. Since models always include $x_1 = e$, so that the coefficient of determination (CD) in regression can be used as a measure of collinearity. Let $\breve{X}_k = \{x_j \in X, j \neq k\}$ in $R^N$ be the set $X$ with the $k$-th variable deleted, where $x_1 = e \in \breve{X}_k$ and $k = 2, \cdots, K$. We also use $\breve{X}_k$ as the $N \times (K-1)$ matrix formed by its columns. Let $\breve{R}_k^2$ is the CD when $x_k$ is regressed on $\breve{X}_k$ in $R^N$. In our terminology, each $x_k$ ($k \geq 2$) is said to be collinear with $\breve{X}_k$ unless $\breve{R}_k^2 = 0$ or $\breve{R}_k^2 = 1$, and it is said to be perfectly collinear with $\breve{X}_k$ if $\breve{R}_k^2 = 1$, and zero-collinear with $\breve{X}_k$ if $\breve{R}_k^2 = 0$. When $\breve{R}_k^2$ is close to 1, $x_k$ is said to be strongly collinear with $\breve{X}_k$. By assuming rank($X$)=$K$, the case of $\breve{R}_k^2 = 1$ is excluded.

In this definition, collinear relations always exist among variables in each model unless $\breve{R}_k^2 = 0$ for any $k \geq 2$, and model performance comes with the collinearity. What to do in modelling is not to avoid it but to exclude each variable $x_k$ that has "strong" collinearity with $\breve{X}_k$ in each model. In the EEM-M, the XMSP will screen such variables out and select models with $\breve{R}_k^2 \leq d$ for any $k \geq 2$, where $d$ is a control parameter.

Concerning collinearity, as its vast literature shows, $y^o$ itself does not have a capacity to identify the collinearity structure of $X$ via a $y^o$MSP. In fact, there will be no literature on such a $y^o$MSP that mitigates the ill-effect within the OLS framework even if $y^o$ is really generated from it, though a lot of hybrid remedies such as shrinkage-type estimation have been provided outside of the OLS framework. Besides, under strong collinearity, $y^o$ loses its own capacity to evaluate model performances by a $y^o$MSP, because OLSEs and maximum likelihood estimates suffer from the ill-effect due to the instability of $(X'X)^{-1}$, which invalidates such a $y^o$MSP as the ones with Akaike's information criterion (AIC), Bayesian information criterion (BIC), adjusted coefficient of determination (CD), etc.

In our terminology, MSP is defined as a procedure to select a model $X^*$ from $X$ that is estimated by the OLS method as $\widehat{\boldsymbol{\beta}}^* = \widehat{\boldsymbol{\beta}}^*(X^*, y)$, where
(1.2) $\quad \widehat{\boldsymbol{\beta}} \equiv \widehat{\boldsymbol{\beta}}(X, y) = (\hat{\beta}_k) = (X'X)^{-1}X'y$.
By the OLSE we mean either OLS estimator $\widehat{\boldsymbol{\beta}}^*$ or OLS estimate $\widehat{\boldsymbol{\beta}}^{*o} = \widehat{\boldsymbol{\beta}}(X^*, y^o)$ for given $(y^o, X^*)$ along its context. Here it is noted that collinearity does not affect the optimality of $\widehat{\boldsymbol{\beta}}$ in (1.2) as a whole. In fact, if the initial model in (1.1) is supposedly "true" with rank($X$) =$K$, $\widehat{\boldsymbol{\beta}}$ is the best linear unbiased estimator in the $K \times K$ nonnegative definite ordering, in which no shrinkage-type estimator will beat the OLS estimator in risk matrix. This optimality holds no matter how strong the collinearity in $X$ may be, implying no collinearity effect on this basic optimality. However, strong collinearity makes variances of some *individual* OLSEs large via the instability of the diagonal elements of $(X'X)^{-1}$ so that some OLSEs in a selected model will be instable in terms of standard errors (SEs) and the empirical effectiveness is lost. Draper and Smith (1998) and Belsley (1991) well describe serious ill-effects on individual estimates with empirical examples and provide some diagnostic methods.

Our results do not much depend on other authors' work. Hence, in the sequel, we summarize our work in some details.

1.1 Empirically Effective Modelling Methodology (EEM-M)

From a practical viewpoint, we first define a new concept of "Empirically Effective Modelling (EEM)" in the OLS framework, where the EEM-M that implements the concept leads to an empirically effective model in its own sense. Here we let $X$ also represents all the cases of sub-model $X^*$.

**Definition 1.1** A model estimated by the OLS with data $(y^o, X)$ is defined to be empirically effective if it is judged to hold the two properties [1] and [2].

[1] Efficiency and stability of each "individual" OLS estimate in the estimated model.

[2] Goodness or optimality in total model performance.

To make this EEM concept implemented, we develop the EEM-M with two MSPs.

The first MSP is the XMSP applied to data $X$ without using $y^o$ for [1] in order to pre-select a class $D$ of inefficiency-controlled and collinearity-controlled models, or shortly "*IC-controlled models*".

In this paper, the measures $(I_k, C_k)$ of inefficiency and instability concerning the property [1] are selected from the individual predictive sampling variance (IndPSV) of each OLS "*estimator*" $\hat{\beta}_k$, which is defined by averaging IndPSVs over $n = 1, \cdots, N$:

(1.3a) $\quad IndPSV_k \equiv \sum_{n=1}^{N} Var(x_{nk}\hat{\beta}_k)/N = \sigma^2 \frac{1}{N}[I_k \times C_k]$,

(see Section 1.3 and (2.3)). Here $I_k$ and $C_k$ are respectively defined for $k \geq 2$ by

(1.3b) $\quad I_k \equiv \bar{x}_k^2/s_k^2 = 1 + \bar{v}_k^2 = 1/(1 - q_k^2) \geq 0$ and

(1.3c) $\quad C_k \equiv VIF_k = 1/(1 - \check{R}_k^2) \geq 0$,

which are respectively called *inefficiency index* $I_k$ and *collinearity index* $C_k$. The variables in (1.3b,c) are given by

$v_{nk} = x_{nk}/s_{xk}$, $\bar{x}_k = \sum_{n=1}^{N} x_{nk}/N$ and $s_{xk}^2 = \sum_{n=1}^{N}(x_{nk} - \bar{x}_k)^2/N$.

Here $q_k^2$ is the squared raw correlation of $x_1 = e$ and $x_k = (x_{1k}, \cdots, x_{Nk})'$ given by

(1.3d) $\quad q_k^2 = (x_k' e)^2/[x_k' x_k \cdot e'e] = cos^2(\theta_k)$,

where $\theta_k$ is the angle of the two vectors. Then, the pair $(I_k, C_k)$ naturally expresses the property [1] of the *k*-th OLS estimate $\hat{\beta}_k$, where both $I_k$ and $C_k$ in (1.3b,c) are free from any measurement unit. For the OLS "*estimate*" $\hat{\beta}_k$ given, the inefficiency index $I_k$ will be called the *I-index* of $\hat{\beta}_k$, while the collinearity index $C_k$ the *C-index* of $\hat{\beta}_k$ ($k \geq 2$).

Now, since values of $(I_k, C_k)$ are determined only by $X$ as in (1.3), the XMSP aims to control the two objective variables

(1.4a) $\quad (I_k(p_k X), C_k(X))$ or equivalently $(q_k^2(p_k X), \check{R}_k^2(X))$

with respect to $X$ in order to obtain a class $D^{cd}$ of "*IC-controlled* models" at control level ($c$, $d$) before $y^o$ is observed, where $p_k X = x_k$ and ($c$,$d$) is prespecified. In other words, we seek such a class of models that satisfy

(1.4b) $\quad I_k \leq c$ and $C_k \leq d$ or equivalently

$$q_k^2 \leq c_q = 1 - 1/c \text{ and } \check{R}_k^2 \leq d_R = 1 - 1/d \text{ for any } k \geq 2.$$

As is well known, the $C_k \equiv VIF_k$ in (1.3c) is the variance inflation factor (VIF) (Jolliffe, 1986), while it is shown in Proposition 2.2 that the $I_k$ index is not only a measure of inefficiency that makes the $k$-th IndPSV large when it is large, but also a measure of collinearity between $x_k$ and $x_1 = e$ as in (1.3d), which is equivalent to a measure of how much $x_k = \{x_{nk}\}$ is immobile over $n = 1, \cdots, N$, because the closer $q_k^2$ is to 1, the closer $x_k$ is to constant vector $e$.

In this paper, as a part of the XMSP, we provide two algorithms to pre-select a class $D^{cd}$ of *IC*-controlled models with (1.4b) in their OLS estimates. Such a class $D^{cd}$ (or simply $D$) will be called an *IC*-controlled class or "$y^o$-accommodating" class for $y^o$MSPs. Once such a class $D^{cd}$ of *IC*-controlled models is obtained, the second MSP in the EEM-M is a $y^o$MSP applied to data $(y^o, X)$ for the property [2] in order to select an optimal model from the class $D^{cd}$. Since there are many useful $y^o$MSPs available, any reasonable $y^o$MSP can be used for finding an optimal model by its own criterion so long as it independently evaluates the total performance of each *IC*-controlled model in $D^{cd}$. In this sense, we will take it for granted that an optimal model can be obtained by such a $y^o$MSP once a target class $D^{cd}$ is obtained. Naturally, the optimality of an *IC*-controlled model depends on the choice of $y^o$MSP, implying that the optimal model is not unique. Examples of $y^o$MSP are those with AIC, BIC, adjusted CD, or hybrids, etc.

However, we will not include the $y^o$MSPs that include a process of pre-testing scheme in selecting models within one sample "$y^o$" (e.g., Saleh, 2006). In view of sampling theory, these MSPs intrinsically entail a nonlinear and conditional structure in a selected model and so it is out of the OLS framework. In addition, those MSPs mostly aim to select significant variables via testing and do not seek an optimal model as a whole.

As has been described above, the XMSP in our EEM-M fills the overlooked gap between (a) the problem that an estimated model needs to have the property [1] for individual OLS estimates and (b) the problem that the traditional $y^o$MSP cannot identify and control the ill-effect of collinearity on individual OLS estimates, because $y^o$ itself have no capacity to do it. In fact, the XMSP fills the weak point of the OLS by preselecting a class $D$ of models with property [1] in advance before $y^o$ is observed, and then a $y^o$MSP comes into own capacity in $D$ to select an optimal model with property [2]. Consequently, the EEM-M will give us an empirically effective model by Definition 1.1.

Another important feature of the EEM-M is that the XMSP developed in Section 3 carries a framework of comparing models in terms of $(I_k, C_k)$. In fact, for comparing models, the maximums $\max_{k \geq 2} I_k$ and $\max_{k \geq 2} C_k$ in a given model $X^*$ are respectively defined to be the inefficiency risk (*I*-risk) and the collinearity risk (*C*-risk) of the model $X^*$, and the pair of the maximums will be defined as the inefficiency-collinearity risk index (ICRI) (see Sections 1.4 and 3.3). This ICRI gives a decision theoretic framework for partial comparison of models because it introduces a partial ordering onto the set of all the sub-models. In particular, in comparing models, they provide the concept of betterness and admissibility of a model in terms of the ICRI.

In addition, as a part of the XMSP, the two algorithms for the XMSP are given for selecting a $D$ of $IC$-controlled models (See Section 4).

These results will suggest that in view of the EEM-M, it is better to have both risk indexes $(I_k, C_k)$ of each $\hat{\beta}_k$ as computer outputs to judge the empirical effectiveness of an estimated model, because these indexes provide the information on how much of the property [1] of each individual OLS estimate is guaranteed whatever $y^o$ may be.

1.2 Bundle Model $\mathcal{BM}([X])$

As a part of the EEM-M, we replace the traditional model concept expressed in (1.1) by our bundle model concept for the EEM-M with the XMSP and $y^o$MSP. This is because the EEM-M starts with observed data $(y^o, X)$ and pursues the effectiveness of empirical model in the sense of Definition 1.1 with some evaluation criteria derived in the OLS framework. As an example, let us consider the problem of selecting one of the models:

(1) $y = X^1\beta^1 + X^2\beta^2 + u$ in (1.1) with $X = (X^1, X^2)$,

(2) $y = X^1\beta^1 + u$.

Such a problem is often treated as a binary decision problem where the $F$-testing scheme is used within the so-called frequentist's framework. This approach leads us to the MSP with a procedure of preliminary test, model selection and estimation. As has been stated, we exclude this MSP because it includes an internal inconsistency so long as the models in the procedure are estimated by the OLS. While, in the EEM-M, the fact that $y$ has been realized as $y^o$ implies that it has to be generated by either (1) or (2). Hence, the error terms in (1) and (2) should be different because the two models cannot generate $y^o$ under the same $u$, unless $X^2\beta^2 = 0$.

Taking this view into the EEM-M framework, we use the bundle model defined by

(1.1a) $\quad \mathcal{BM}([X]) = \{ y_\tau = X_\tau \beta_\tau + u_\tau \mid X_\tau \in [X],\ u_\tau \in [u] \}$.

Here $[X]$ is the set of all the sub-models $\{X_\tau : \tau \in \Lambda\}$ with $x_{\tau 1} = e$ and $\Lambda = \{1, 2, \cdots, 2^{K-1} - 1\}$, where $\tau$ is a parameter that distinguishes models, and $[u]$ is the set of error terms corresponding to the set $[X]$:

(1.1b) $\quad [u] = \{u_\tau \mid \tau \in \Lambda,\ E(u_\tau) = 0, Var(u_\tau) = \sigma^2 I\}$.

Then $y^o$ is regarded as realized from one of the sub-models in $\mathcal{BM}([X])$, not from a "true model". It is noted that no specific stochastic structure is specified among $u_\tau$'s except for its own two moments as in (1.1b). This is because any $y^o$MSP in the EEM-M is assumed to be able to evaluate the total performance of each individual model in $B\mathcal{M}([X])$ with its own criterion when $(y^o, X)$ is given. In Section 3.2, a further discussion is made on this point in view of model comparison.

In the EEM-M, the XMSP will replace $[X]$ in $B\mathcal{M}([X])$ by a class $D$ of *IC*-controlled models, which makes $y^o$MSP better accommodated or equivalently more effectively functioned. This can be done in advance before $y^o$ is observed.

The above argument shows that in our analytical framework, there is no concept of "true" model. After all, a finally selected empirically effective model via a $y^o$MSP will have to be regarded as the model having generated $y^o$ in the EEM-M or even in any empirical analysis, so long as $y^o$ is regressed on the final model and it is used in applications.

1.3 Individual Predictive Sampling Variance for *I*-index and *C*-index

To describe the two indexes in (1.3) for individual efficiency of OLSEs in a sub-model $y^* = X^*\beta^* + u^*$, let us first consider the individual sampling variance (IndSV) $Var(\hat{\beta}_k^*)$ of each OLSE $\hat{\beta}_k^*$. Then, as is well known, it is decomposed as

(1.5) $\quad Var(\hat{\beta}_k^*) = \dfrac{\sigma^2}{EEF_k^2} = \dfrac{\sigma^2}{Ns_{xk}^2} \times VIF_k \quad$ with $\quad EEF_k^2 = Ns_{xk}^2(1 - \breve{R}_k^2),$

where $k = 2, \cdots, K^*$. Here $EEF_k^{-2}$ is the (*k,k*) element of $(X^{*\prime}X^*)^{-1}$ and $EEF_k$ is called empirically effective factor (EEF) that controls $Var(\hat{\beta}_k^*)$. To make $Var(\hat{\beta}_k^*)$ smaller, $EEF_k$ is desired to be larger. Also, $Var(\hat{\beta}_k^*)$ in (1.5) is linearly affected by *C*-index $C_k \equiv VIF_k$ in (1.3b).

Clearly each $EEF_k^2$ consists of the two components:
1) the variate-own effect $Ns_{xk}^2$ of predictive variable $x_k^*$ and
2) the collinearity effect $C_k$.

Hence, for given $C_k$, the larger the $Ns_{xk}^2$ is, the smaller $Var(\hat{\beta}_k^*)$ is, and the more efficient the *k*-th OLSE is. This is inappropriate in comparing the standard errors (SEs) $[\widehat{Var}(\hat{\beta}_k^*)]^{1/2}$'s of individual OLSEs in an estimated model. Besides, the physical units of $x_k^*$'s in measurement are different in general and so IndSVs in (1.5) are not comparable.

To overcome the incomparability of IndSVs, we pay attention to individual terms $x_{nk}\hat{\beta}_k^{*\prime}s$ in each model. Since these terms have a common physical unit with $y_n$, we adopt the predictive sampling variance (IndPSV) of each $\hat{\beta}_k^*$ that is defined by the average of $Var(x_{nk}\hat{\beta}_k)$ over $n = 1, \cdots, N$ as in (1.3a), which is re-expressed as

(1.6) $\quad IndPSV_k = \left[\dfrac{\sum_{n=1}^{N} x_{nk}^2}{N}\right] Var(\hat{\beta}_k^*) = \sigma^2 \dfrac{1}{N} [I_k \times C_k]$

$$= \sigma^2 \frac{1}{N}\left(\frac{1}{1-q_k^2}\right)\left(\frac{1}{1-\check{R}_k^2}\right) \equiv \sigma^2 \frac{1}{N}\widetilde{H}(q_k^2, \check{R}_k^2)$$

(see (2.3)). Then it is clear that the two indexes $I_k$ and $C_k$ in (1.3b,c) are free from measurement unit since they are scale-invariant (see Section 3). Note that $\sigma$ carries the same physical unit as $y_n$. And the $IndPSV_k$ of the $k$-th term $\{x_{nk}\hat{\beta}_k^*: n = 1, \cdots, N\}$ consists of the two effects:

3) the variate-own inefficiency effect $I_k$ and

4) the collinearity effect $C_k$.

It is desirable for both $I_k(\geq 1)$ and $C_k(\geq 1)$ to be smaller so that $IndPSV_k$ in (1.6) becomes smaller. In (1.6), for the $k$-th OLS estimate in model $X^*$, $I_k$ is the $I$-index to be controlled for efficiency as in (1.3b), while $C_k$ is the $C$-index to be controlled for stability (see (1.3c)).

In Section 2, for reference, the standard error (SE) of each estimate based on IndSV is compared to the SE based on IndPSV after model is estimated with $y^o$. Also, we give a necessary and sufficient condition for a model $X^*$ to attain the lower bound for the IndPSVs of all the estimates.

1.4 $y^o$-Accommodating Class, IC Risk Index and Admissibility of Model

In Section 3, we will formulate a decision theoretic framework for comparing models by the IC Risk Index of $X^*$. Since the degree of collinearity of model $X^*$ is not greater than that of $X^{**}$ if $X^* \subset X^{**}$, it is necessary to take the column size into account in comparing models by $VIF_k$. Lavery, Acharya, Sivo and Xu (2019) studied on the column size and collinearity via simulations with focus on errors and bias. Here, when models are compared in terms of collinearity index, the column size effect is taken into account. Let $J(X^*)$ denote the column size of $X^*$. Then, the distinguishability of models $\{X^*\}$ with $J(X^*) = J$ can be made only through the set of ($I$, $C$)-indexes of individual OLS estimates in each $X^*$;

(1.7) $\quad \varphi(X^*) \equiv \{(I_k(p_k X^*), C_k(X^*)) | k = 2, \cdots, J\}$,

where $p_k X^* = x_k^*$. And taking the column size effect on $C$-index into consideration, let $D_J^{cd}$ denote the class of $IC$-controlled models of column size $J$ with control level ($c$, $d$), where

(1.8a) $\quad D_J^{cd} = \{X^* \in [X:J] | \ I_k(p_k X^*) \leq c, \ C_k(X^*) \leq d, \ (k = 2, \cdots, J)\}$,

where $[X:J]$ denotes the set of models whose column sizes are $J$. Also let

(1.8b) $\quad\quad D^{cd} = \cup_{J=2}^{K} D_J^{cd}$

be the set of all the $IC$-controlled models with control level ($c$, $d$). The problem of selecting (*c*, *d*) is treated in Section 3.2. To materialize the XMSP, two algorithms are proposed for finding $D^{cd}$ in Section 4.

Also in Section 3, a partial ordering is introduced onto the set of models of column size *J* based on the IC Risk Index of each model $X^*_{\tau J}$, which is defined by

(1.9a) $\quad r(X^*_{\tau J}) = (c_{M\tau J}, d_{M\tau J})$ for each $X^*_{\tau J}$,

where the inefficiency risk and the collinearity risk of $X^*_{\tau J}$ are respectively defined by

(1.9b) $\quad c_{M\tau J} = \max_k I_k(p_k X^*_{\tau J})$ and $d_{M\tau J} = \max_k C_k(X^*_{\tau J})$.

The model $X^*_{\tau J}$ with IC risk index in (1.9b) is also denoted as $X^*_{\tau}(c_{M\tau J}, d_{M\tau J})$. Then for each *J* fixed, $X^*_{\tau}(c_{M\tau J}, d_{M\tau J})$ is said to better accommodate $y^o$ than $X^*_{\tau'}(c'_{M\tau J}, d'_{M\tau J})$ if $c_{M\tau J} \leq c'_{M\tau J}$ and $d_{M\tau J} \leq d'_{M\tau J}$ hold with one of the inequalities strict. Importantly, this can be identified before $y^o$ is observed. From this set-up, the concept of admissibility of model naturally follows (see Section 3), though the characterization of admissible $y^o$-accommodating class of models is made only in the case of *J*=2.

1.5 Two Algorithms for Materializing the XMSP and a Numerical Example

In Section 4, to make the XMSP practically feasible, we develop two computational algorithms: variable-increasing and variable-reducing algorithms. In the latter case, principal component analysis (PCA) is used, and it is shown that $\check{R}^2_k(X) = \check{R}^2_k(Z)$ for the matrix *Z* of standardized variates of $x_j$ ($k \geq 2$). In particular, we focus on the algorithm to make models satisfy the condition $\check{R}^2_k \leq d_R \equiv 1 - 1/d$ for any *k* after variables that satisfy $q^2_k \leq c_q = 1 - 1/c$ are found (see (1.4c)). Then a set $\{X^*\}$ of *IC*-controlled models is obtained by combining them as the intersection of these sets. As a numerical example, applying the algorithms to the gasoline consumption data given in Chatterjee and Hadi (2012), we will find a class of *IC*-controlled models in the cases of real data and simulation data.

1.6 A brief Literature Review on Collinearity in Model Selection

Research history on collinearity in regression is very long and has been still accumulating a vast amount of literature, though no clear-cut solution exists. In our OLS context, a recent example is Tsao (2019) proposing estimating a linear combination of regression parameters in a strongly correlated model *X*. While, because they are not in the OLS framework, we do not treat such hybrid methods and procedures of the GLS-type (Kariya and Kurata, 2003), LASSO-type (Zou and Hastie, 2005), ridge-type (Hoerl and Kennard, 1988), Stein-type (Kubokawa and Srivastava, 2004), and principal component-type (Jolliffe, 1986) methods.

These sophisticated methods basically aim to secure the stability of estimates in the context of collinearity. Stewart (1987) mathematically clarified the algebraic structure of collinearity in $X$. Fox and Monette (1991) generalized the concept of VIF to the case where $X$ is divided into three categories of variables, including dummy variables. Lavery, Acharya, Sivo and Xu (2019) studied on the relation between the number of variables and collinearity on errors and bias. Aitken and West (1996) showed in a three-variate linear model with interaction term that standardizing variables $(x_{nk} - \bar{x}_k)/s_{xk}$ in modelling significantly reduces the collinearity among the variables.

Some $y^o$ MSPs are associated with a pre-testing, variable selection and estimation procedure and they are often used as MSPs of stepwise forward, backward or hybrid methods. In their textbook, Draper and Smith (1998) well describe them with various empirical examples. But these MSPs do not lead to an optimal model as a whole. In addition, the ill-effect of collinearity may let us select a wrong model. Chatterjee and Hadi (2012) also heuristically exposit some details of empirical regression analysis with real data, which includes collinearity cases in data.

2  Sampling Variance and Predictive Sampling Variance of OLSE

In this section, using the individual predictive sampling variance (IndPSV) of each individual OLS estimate in (1.6), we study some important relations between IC indexes $(I_k, C_k)$ in (1.3b,c) and the structure of model $X^*$ that carries $\{(I_k, C_k)|k = 2, \cdots, K^*\}$. Then, we compare the traditional standard errors (SEs) of individual OLS estimates based on the IndSVs and the corresponding predictive standard errors (PSEs) based on the IndPSVs. In the sequel, we let $X$ represent all the cases of $X^* \in [X]$.

The effectiveness of the XMSP relies on the fact that the IndPSV of each term $\{x_{nk}\hat{\beta}_k\} = \{x_{nk}\hat{\beta}_k|n = 1, \cdots, N\}$ with $\hat{\beta}_k$ being the OLS "estimator" is expressed as $\sigma^2 \widetilde{H}(q_k^2, \check{R}_k^2)/N$ in (1.6), which is a function of $q_k^2$ in $I$-index $I_k \equiv 1/(1 - q_k^2)$ and $\check{R}_k^2$ in $C$-index $C_k \equiv 1/(1 - \check{R}_k^2)$. Since the $\widetilde{H}$ function depends only on $X$, an "estimated" model can be made $IC$-controlled by controlling the $I$-index and the $C$-index separately. Hence, the IndPSVs as a sampling property of $\{x_{nk}\hat{\beta}_k\}$ is connected with the effectiveness of the OLS estimates in the estimated model before $y^o$ is observed. It is noted that the sample size $N$ is fixed throughout the paper.

2.1 Relations between $(I_k, C_k)$ and the Structure of $X$

Now, let us consider an estimated predictive model:
(2.1)    $\hat{y}_n = \hat{\beta}_1 + x_{n2}\hat{\beta}_2 + \cdots + x_{nK}\hat{\beta}_K$,
which is the $n$-th element $\hat{y}_n$ of $\hat{y} = \sum_{k=1}^{K} x_k \hat{\beta}_k = X\hat{\beta}$. This is also viewed as pre-sampled version as $\hat{y}_n \equiv \hat{y}_n(y)$ for given $X$. Then, for $\{x_{nk}\}$ given, the sampling distribution of $\hat{y}_n(y)$ is dependent on the whole covariance matrix $XVar(\hat{\boldsymbol{\beta}})X'$, but in view of the EEM, what matters in the post-sampled predictive model with $\boldsymbol{y}^o$ given is the efficiency and stability of individual terms $\{x_{nk}\hat{\beta}_k\}'s$ by Definition 1.1. Then, from the above observation, it suffices to control the inefficiency index (*I*-index) and the collinearity index (*C*-index) of each term in (2.1), since it controls $Var(x_{nk}\hat{\beta}_k)$ for each $x_{nk}$ whether or not $\boldsymbol{y}^o$ is observed. If the IndPSV is small, the contribution of $x_{nk}\hat{\beta}_k$ to $\hat{y}_n$ will be stable and $x_{nk}\hat{\beta}_k$ is likely to be realized in a neighborhood of its mean $x_{nk}\beta_k$ for fixed $x_{nk}$, though the variation is that of $\hat{\beta}_k$. Since $x_{nk}$ varies over $\{x_{nk}\}$, averaging $Var(x_{nk}\hat{\beta}_k)$ over $n = 1, \cdots, N$ yields the IndPSV as

$$(2.2) \quad IndPSV_k \equiv \frac{1}{N}\sum_{n=1}^{N} Var(x_{nk}\hat{\beta}_k) = [\sum_{n=1}^{N} \frac{x_{nk}^2}{N}]Var(\hat{\beta}_k)$$

$$= \sigma^2 [\sum_{n=1}^{N} \frac{x_{nk}^2}{N}] \frac{1}{Ns_{xk}^2(1-\check{R}_k^2)} \equiv \sigma^2 \frac{1}{N} H_k.$$

Hence, letting $v_{nk} = x_{nk}/s_{xk}$ and using

$$\sum_{n=1}^{N} x_{nk}^2 / Ns_{xk}^2 = 1 + \bar{v}_k^2 = 1/(1 - q_k^2) \equiv I_k,$$

$H_k$ in (2.2) is expressed as

$$(2.3) \quad H_k = H(I_k, C_k) \equiv I_k \cdot C_k = \left(\frac{1}{1-q_k^2}\right)\left(\frac{1}{1-\check{R}_k^2}\right) \equiv \widetilde{H}(q_k, \check{R}_k^2).$$

It is noted that when $I_k \leq c$ and $C_k \leq d$ for any $k$,

$$(2.4) \quad H_k \leq \Delta_{cd} \equiv cd \text{ for any } k.$$

This $H_k$ is clearly free from the physical unit of $x_{nk}$ so that its size is comparable with those of the other $H_j's$ $(j \neq k)$ in each model, and the physical unit of $y$ is carried over to its population standard deviation $\sigma$ in (2.2). In (2.3), the IndPSV is separated into the inefficiency index $I_k$ and the collinearity index $C_k$.

It is remarked that though $N$ is fixed throughout this paper, taking the average in (2.2) guarantees that, when

$(2.5) \quad lim_{N\to\infty}(X'X/N) = A$ exists (finite) with $|A| \neq 0$,

$IndPSV_k$ converges to 0 as $N \to \infty$ because $Var(\hat{\beta}_k)$ converges to 0 under (2.5). In fact, $\bar{v}_k^2 = \bar{x}_k^2/s_{xk}^2$ and $C_k$ respectively converge to corresponding constants. The condition (2.5) is often and typically assumed in developing asymptotic argument in regression.

Now let us study the structure of model $X$ for which each individual OLSE satisfies the efficiency and stability condition [1] in Definition 1.1. From (2.3), it is easy to observe the following facts. Letting " $\Leftrightarrow$ " be read as "if and only if",

(a) $\bar{v}_k^2 = 0 \Leftrightarrow \bar{x}_k = 0 \Leftrightarrow q_k^2 = 0$, and $C_k = 1 \Leftrightarrow \check{R}_k^2 = 0$.

(b) $H_k = 1 \Leftrightarrow [\bar{x}_k = 0$ and $\check{R}_k^2 = 0$ ].

When $\bar{x}_k = 0$, then $H_k = C_k$, and then controlling $C_k$ is equivalent to controlling the IndPSV in (2.2). While, when $C_k = 1$, then $H_k = I_k$ so that it is desirable for $I_k$ to be small. When $H_k (\geq 1)$ is close to 1, the term $\{x_{nk}\hat{\beta}_k\}$ in (2.1) is stabilized in the sense of $IndPSV_k$ and the collinearity index is not large whether or not $y$ is observed.

The next proposition gives a necessary and sufficient condition for the attainment of the lower bound of $H_k$. The proof is given in Appendix 1.

Proposition 2.1. In (2.3), for given $k(\geq 2)$, $H_k = 1$ if and only if $\bar{x}_k = 0$ and $x_i' x_k = 0$ for any $i \neq k$ ($i \geq 1$). Hence, $H_k = 1$ for any $k(\geq 2)$ if and only if

(1) $\bar{x}_k = 0$ for any $k(\geq 2)$ and (2) $X'X = diag\{a_1, a_2, \cdots, a_K\}$,

where $a_1 = N$ and $a_i = x_i' x_i$. If model satisfies (2) only, $H_k = I_k = 1/(1 - q_k^2)$ only denotes the effect of inefficiency. Here $diag\{a_1, \cdots, a_K\}$ denotes diagonal matrix.

Example 2.1 An example of $X$ satisfying the conditions (1) and (2) in Proposition 2.1 is $X = \{x_1, x_2, \cdots, x_K\}$ with $x_i = \alpha_i \delta_i$ ($\alpha_i > 0$), where $E \equiv (\delta_1, \delta_2, \cdots, \delta_K): N \times K$ is the matrix consisting of the first $K$ columns of the well-known Helmert's orthogonal matrix with

$$\delta_1 = e/\sqrt{N}, \quad x_2 = (\frac{1}{\sqrt{2 \cdot 1}}, \frac{-1}{\sqrt{2 \cdot 1}}, 0, \cdots, 0)',$$

$$\delta_k = (\frac{1}{\sqrt{k \cdot (k-1)}}, \frac{1}{\sqrt{k \cdot (k-1)}}, \cdots, \frac{1}{\sqrt{k \cdot (k-1)}}, \frac{-(k-1)}{\sqrt{k \cdot (k-1)}}, 0, \cdots, 0)' \text{ for } k \in \{2, \cdots, N\}.$$

In fact, this matrix satisfies $\bar{x}_k = 0$ and $X'X = diag\{a_1, a_2, \cdots, a_K\}$ with $a_i = \alpha_i^2$. This fact gives an interesting implication to the Helmert's orthogonal matrix. On the other hand, since it is assumed that model (1.1) includes constant term $x_1 = e$, $X' = [I, 0']: K \times N$ cannot be a model for $X$ satisfying (1) and (2).

Next, let us study some properties of the *I*-index, which will be used in Section 3.

Proposition 2.2 (1) When $K \geq 3$, the inefficiency index is measured independently of the collinearity index $VIF_k$ and $I_k = 1 + \bar{v}_k^2$ is expressed as

(2.6) $\quad 1 + \bar{v}_k^2 = 1 + \frac{\bar{x}_k^2}{s_k^2} = 1 + \frac{x_k' M_e x_k}{x_k'(I - M_e)x_k} = 1 + \frac{q_k^2}{1 - q_k^2} = \frac{1}{1 - q_k^2},$

which takes values on [1, ∞) and is increasing in $q_k^2 \in [0,1)$, where

$$q_k^2 = (x_k' e)^2 / [x_k' x_k \cdot e' e] = cos^2(\theta_k)$$

is the squared raw correlation of $x_1 = e$ and $x_k$, and $M_e = e(e'e)^{-1}e'$.

(2) If $K=2$, then $I_2 = 1/(1 - q_2^2)$ and $C_2 = 1/(1 - q_2^2)$, and hence the two indexes are

both a function of $q_2^2$ and $H_k = 1/(1-q_2^2)^2$ holds. Therefore, they are not separable, and $H_k = 1$ if and only if $x_k'e = 0$.

(3) For $K \geq 2$, $I_k = 1/(1-q_k^2)$ is a collinearity measure between $x_1 = e$ and $x_k$, which is also a measure of immobility of $x_k = \{x_{nk}\}$ over $n = 1, \cdots N$.

Proof. Straightforward and omitted.

Thus, by Proposition 2.2 (1)(3), the closer the angle between two vectors $x_k$ and $e$ is to 0, the larger the $I$-index $I_k$ is. In other words, if and only if $x_k$ and $e$ are orthogonal, $I_k$ attains the minimum 1, implying that $I_k$ is also measuring the collinearity of $x_k$ with $e$. Also, because large $I_k$ means that $q_k^2$ is close to 1, it is a measure of immobility of $x_k = \{x_{nk}\}$ over $n = 1, \cdots N$. Hence, it will be better to choose such variables $x_k$'s that $cos^2(\theta_k)$ is close to 0 or equivalently $x_k$ is less collinear with $e$ so that the $I$-index is smaller, provided such a selection of variables is possible.

2.2 Comparisons of IndSV $Var(\hat{\beta}_k)$ and IndPSV $IndPSV_k$

To compare IndSV and IndPSV, let us make some basic observations.

First, in the case of IndSV $Var(\hat{\beta}_k)$ in (1.5) that depends on $(s_{xk}, VIF_k)$, it is desirable for $s_{xk}$ to be large in order to make the IndSV small. While, in the case of the IndPSV that depends on $(I_k, C_k)$ in (2.2), $I_k = 1 + \bar{v}_k^2$ is desired to be small, where $\{v_{nk} = x_{nk}/s_{xk}\}$ satisfies $s_{vk}^2 = \sum_{n=1}^N (v_{nk} - \bar{v}_k)^2/N = 1$. Hence, under $s_{vk} = 1$, it is desirable for $(\bar{x}_k)^2$ to be small as well as for $s_{xk}$ to be large for the sake of the IndSV. Consequently, so long as $I_k$ is concerned, it will be better to have variable $\{x_{nk}\}$ distributed over a broad interval including 0 with its mean close to 0 and standard deviation $s_{xk}$ large, so that $\bar{v}_k^2$ becomes small. Then such a variable will be better for IndSV as well as IndPSV, though it may not be found in $X^*$ given. Hence, if $s_{xk}$ is also desired to be large for the sake of IndSV as in (1.5), then the distribution of $\{x_{nk}\}$ needs to be distributed more like $N(0, \gamma)$ with large $\gamma$. On the other hand, the case of $\{x_{nk} > 0$ for any $n\}$ is likely to make the $I$-index $I_k$ large on $[0, \infty)$ (see (2.6)), which is controlled by the XMSP. In other words, to avoid inefficiency, the XMSP will select such model that $x_k$ satisfies $I_k \leq c$ for any $k$, where $y^o$ is not used. In addition, Proposition 2.2 is used to make a reasonable choice of the threshold value $c$ for $I_k$, because $I_k \leq c$ is equivalent to $q_k^2 \leq c_q = 1 - 1/c$.

Second, if standardized variable $w_{nk} = (x_{nk} - \bar{x}_k)/s_{xk}$ can be used for $x_{nk}$ in the model, then $IndPSV_k = \sigma^2 C_k/N$ because $\bar{w}_k = 0$. This is an important implication for empirical modelling in treating collinearity via standardization, *when the IndPSV is used*. In Section 4, it will be shown that the C-Index based on $\{w_{nk}\}$ is the same as the one based on $\{x_{nk}\}$.

Third, the XMSP does not aim to minimize $H_k's$ nor its mean, but aims to control the two indexes of $I_k$ and $C_k$ separately because the strong collinearity of $x_k$ can seriously cause confounding effects on the other variables. Though, $H_k \leq cd$ holds as in (2.4).

Now, let us compare the standard error (SE) of $\hat{\beta}_k$ and the corresponding predictive standard error (PSE), which are respectively given by

(2.7a) $\quad SE\left(\hat{\beta}_k(X^*)\right) = \hat{\sigma}(y^o, X^*)(Ns_{xk}^2)^{-\frac{1}{2}}\sqrt{C_k}$, and

(2.7b) $\quad PSE\left(\hat{\beta}_k(X^*)\right) = \hat{\sigma}(y^o, X^*)\frac{1}{\sqrt{N}}\sqrt{I_k}\sqrt{C_k} \equiv \hat{\sigma}\frac{1}{\sqrt{N}}H_k^{1/2}$,

where $\hat{\sigma}$ is the residual standard error (RSE) under model $X^*$ via OLS with $y^o$, which is another important measure of model performance with $y^o$. Each IndSV in (2.7a) can be smaller than $\sigma^2$ because of $(Ns_{xk}^2)^{-1/2}$ with large $s_{xk}$. Hence, it is not easy to compare and interpret those SEs, although the SEs in (2.7a) have been traditionally given as computer outputs with other statistics such as adjusted CD (coefficient of determination). In the case of $\hat{\sigma}H_k^{1/2}/\sqrt{N}$ in (2.7b), which we call the PSE of the term $\{x_{nk}\hat{\beta}_k : n = 1, \cdots, N\}$, each $H_k^{1/2}$ can be compared with other $H_i^{1/2}$'s. Hence the individual estimates in (2.2) are comparable in terms of the PSEs together with RSE $\hat{\sigma}$. In empirical analysis, it will be better to have the PSE values $PSE(\hat{\beta}_k(X^*))'s$ as computer outputs that reveal the contribution of each term $\{x_{nk}\hat{\beta}_k\}$ to $y^o$ with relative roles of $x_k^*\beta_k's$.

Finally it is remarked that by (1.5), the $k$-th SE is expressed as

(2.8) $\quad SE\left(\hat{\beta}_k(X^*)\right) = \hat{\sigma}(y^o, X^*)\|x_k^*\|^{-1}\frac{1}{\sqrt{N}}\sqrt{I_k}\sqrt{C_k}$.

where $\|x_k^*\|^2 = \sum_{n=1}^N x_{nk}^2$. This expression shows that the $k$-th SE depends on the three effects; $\|x_k^*\|$, $I_k$ and $C_k$. In Section 3, it will be pointed out from (2.8) that the SE is controlled by controlling $\|x_k^*\|$ *after* a $y^o$-accommodating class of models with the $I_k$ and $C_k$ controlled is obtained, where the three effects are functionally independent if $K^* \geq 3$.

As was stated, the performances of different models $X^{*'}s$ may be compared by the mean of $\{H_k\}$ as

(2.9) $\quad \bar{H} \equiv \bar{H}(X^*) = \frac{1}{K^*-1}\sum_{k=2}^{K^*} H_k = \frac{1}{K^*-1}\sum_{k=2}^{K^*} I_k \cdot C_k \leq cd$

for $X^*: N \times K^*$ in the IndPSVs, provided $I_k's$ and $C_k's$ are controlled in such a way that $I_k's \leq c$ and $C_k \leq d$ for any $k$. This gives a direct comparability of different models $\{X^*\}$ in a certain class of such models as, e.g., those of the same column size $J$ as in Section 3.3. However, in model comparison, it will be necessary to take the RSE $\hat{\sigma}(\boldsymbol{y}^o, \boldsymbol{X}^*)$ into consideration, implying that after a class of *IC*-controlled models is obtained, models in the class may be compared in terms of $\hat{\sigma}(\boldsymbol{y}^o, \boldsymbol{X}^*)\overline{H}(\boldsymbol{X}^*)/\sqrt{N}$ for each fixed $J$. In addition, it will be necessary to use a $\boldsymbol{y}^o$MSP to evaluate a whole model performance with $(\boldsymbol{y}^o, \boldsymbol{X})$ and select an optimal model by such MSPs with AIC, BIC, adjusted CD $\overline{R}^2$, etc.

## 3  The XMSP for EEM-M

In this section we implement the XMSP together with the algorithms in Section 4. First of all, recall that the whole EEM-M process is symbolically described as $\mathcal{BM}([\boldsymbol{X}]) - \text{XMSP} - \boldsymbol{y}^o\text{MSP}$ (see Section 1.2 for $\mathcal{BM}([\boldsymbol{X}])$). In particular, the XMSP in the EEM-M aims to control $(I_k, C_k)$ in (1.3) in such a way that $I_k \leq c$ and $C_k \leq d$, before $\boldsymbol{y}^o$ is observed. Hence, unlike diagnostic checking procedure after a model is estimated, the XMSP enables us to pursue an empirically effective model by finding a class $D$ of *IC*-controlled models for efficiency and stability of OLS estimates. The class is also called a $\boldsymbol{y}^o$-accommodating class for $\boldsymbol{y}^o$MSP.

In Section 3.1, the bundle model $\mathcal{BM}([\boldsymbol{X}])$ in Section 1.2 is further discussed here to develop the XMSP and in Section 3.2, we discuss some methods of obtaining a class $D$ of *IC*-controlled models, where the column size of each model $\boldsymbol{X}^*$ is taken into account. In Section 3.3, for each model $\boldsymbol{X}^*$, using inefficiency-collinearity risk index (ICRI) (see Section 1.4), a partial ordering is introduced onto the set $[\boldsymbol{X}]$ to compare models in terms of ICRI and then the concept of admissibility of model is defined based on the ICRI.

### 3.1 Bundle Model and Stochastic Specification

Recall that the bundle model in (1.1a, b) is given by

(1.1a)  $\mathcal{BM}([\boldsymbol{X}]) = \{\boldsymbol{y}_\tau = \boldsymbol{X}_\tau \boldsymbol{\beta}_\tau + \boldsymbol{u}_\tau \mid \boldsymbol{X}_\tau \in [\boldsymbol{X}],\ \boldsymbol{u}_\tau \in [\boldsymbol{u}]\}$ with

(1.1b)  $[\boldsymbol{u}] = \{\boldsymbol{u}_\tau \mid \tau \in \Lambda,\ E(\boldsymbol{u}_\tau) = \boldsymbol{0}, Var(\boldsymbol{u}_\tau) = \sigma^2 \boldsymbol{I}\}.$

This is the set of different models with their own error terms from which we aim to select a model to use for empirical applications in view of the EEM-M. Note that $\boldsymbol{y}^o$ is regarded as generated from a model in the bundle model. This model implies $Var(\boldsymbol{y}_\tau) = \sigma^2 \boldsymbol{I}$ for *any sub-model* $\boldsymbol{X}_\tau$ in the bundle set, which in turn implies the common structure of $H(I_{\tau k}, C_{\tau k})$ in (2.3) for any $\boldsymbol{X}_\tau$. This is the objective function to be controlled to select a $\boldsymbol{y}^o$-accommodating class $D$ in the XMSP and it is symbolically expressed: for $\tau \in \Lambda$,

$$\vartheta_\tau \equiv (\{H(I_{\tau k}, C_{\tau k}) \mid k = 2, \cdots, K_\tau\}, Var(\boldsymbol{y}_\tau) = \sigma^2 \boldsymbol{I}),$$

which is independent of $\boldsymbol{y}_\tau$ (including $\boldsymbol{y}^o$) but dependent on $\boldsymbol{X}_\tau$. This structure is in fact an essential part of empirical modelling connected with the predictive sampling variances of individual OLS *estimators* and it enables the XMSP to aim to select a class of *IC*- controlled models by controlling all the $(I_{\tau k}, C_{\tau k})'s$ of individual OLS *estimates* in each model without observing $y^o$. Here we drop the suffix $\tau$ distinguishing models unless it is necessary and denote a representative sub-model by $X^*$.

To develop the XMSP, first note that when model sequence $\{X_i\}$ is increasing in such a way as $X_i \subset X_{i+1} \subset \cdots \subset X$, then the $k$-th CD $\check{R}_k^2(X_i)$ or equivalently $VIF_k$ is increasing in the column size of each matrix $i$ for fixed $k$. Hence let

(3.1a) $\quad [X] = \cup_{J=2}^{K} [X:J]$ with

(3.1b) $\quad [X:J] = \{X^* \in [X] | X^* : N \times J, \ rank(X^*) = J, \ x_1 = e \in X^*\}$.

Note $[X] \equiv [X:K]$. To connect $X^* \in [X]$ with $y$, let $u^* \in [u]$ represent a corresponding error term that has no association with $(y, X)$ when $X^* \neq X$. Then the observed $y^o$ is regarded as realized through one of the models in the following set:

(3.2a) $\quad BM([X]) = \cup_{J=2}^{K} BM([X:J])$ with

(3.2b) $\quad BM([X:J]) = \{y^* = X^* \beta^* + u^* | \ X^* \in [X:J], \ u^* \in [u] \}$.

The model (1.1) itself belongs to $BM([X])$.

3.2 $y^o$-Accommodating Class : Controlling $(I_k, C_k)$

In the bundle model (3.2), the XMSP is a process of exploring a class of models $\{X^*\}$ with $I_k's$ and $C_k's$ controlled separately, so that all the IndPSVs of each model in the selected class are better controlled. This implies that all the models in the class are candidates of empirically effective models because an optimal model is selected from them by a $y^o$MSP. Of course, the optimality depends on the choice of the $y^o$MSP. In other words, the XMSP is a self-fulfilling process to make a selected class of models *IC*-controlled for $y^o$MSPs.

Now, we formally define the concept of $y^o$-accommodating set of models. Let $J(X^*)$ denote the column size of matrix $X^*$, and let $p_k X^* = x_k^*$ for $k \leq J(X^*)$ as before.

Definition 3.1 Let $(c, d)$ be prespecified as control parameters. The $y^o$-accommodating class of models at level $(c, d)$ is defined to be the class $D^{cd} \equiv D^{cd}([X])$ of models, where

(3.3) $\quad D^{cd} = \{X^* \in [X] | \ I_k(p_k X^*) \leq c, \ C_k(X^*) \leq d, \ (k = 2, \cdots, J(X^*))\}$
$\quad \quad \quad = D_1^c \cap D_2^d$.

Here

(3.3a) $\quad D_1^c = \{X^* \in [X] | \ I_k(p_k X^*) \leq c, \ (k = 2, \cdots, J(X^*))\}$, and

(3.3b) $\quad D_2^d = \{X^* \in [X] | \ C_k(X^*) \leq d, \ (k = 2, \cdots, J(X^*))\}$.

Proposition 3.1. (1) If $X^* \in D^{cd}$ and $X^{**} \subset X^*$, then $X^{**} \in D^{cd}$.
(2) If $X_1, X_2 \in D^{cd}$, then $X_1 \cap X_2 \in D^{cd}$.

This obvious proposition is practically useful to find a model in $D^{cd}$. If $X^*$ belongs to $D^{cd}$, so does any sub-model $X^{**}$ of $X^*$. This will be used in Section 4 to develop an algorithm of finding models in $D^{cd}$ via PCA.

Now, to get a $D^{cd}$, we need to choose $(c, d)$, but it will be difficult to discuss the choice in a unified manner. A difficulty lies in the fact that the set of variables in data $X$ is given. In many areas of social sciences, we have to use the data as it stands. Then, so long as the systems generating $X$ are not so volatile, $\check{R}_k^{2\prime}s$ will carry some predictability as a whole for dependent variable. On the other hand, in natural sciences where data $X$ can be designed to a large extent, then $\check{R}_k^{2\prime}s$ will be controlled with smaller $d$ and then $I_k's$ will be more concerned. In general, it will be reasonable to first find $D_1^c$ by deleting variables with $I_k's$ larger than $c = /(1 - c_q)$, or equivalently $q_k^{2\prime}s \leq c_q = 1 - 1/c$, and then to find $D_2^d$ of models with $\check{R}_k^{2\prime}s \leq d_R = 1 - d^{-1}$ from $D_1^c$. This is because $q_k^2$ depends only on its own variable and because the effect of collinearity on variables is confounding and susceptible to addition or deletion of variables. And so it is better to first get the set of variables via $q_k^{2\prime}s$, and then the effect of collinearity is considered via $\check{R}_k^{2\prime}s$. Besides, it will save some computation.

Some remarks follow. First, in the literature, concerning collinearity, it is often suggested that in an MSP via VIF, $x_k^*$ be dropped if $VIF_k > 10$ (or 5) or equivalently $\check{R}_k^2 > 0.9$ (or 0.8 respectively) in each stepwise model selection with $y^o$, though there is no theoretically solid ground (e.g., O'Brien (2007)) and it is likely to lead to a wrong model since the final model is dependent on the order of selection. In addition, there is no control on the $I_k$-index.

Second, once $D^{cd}$ is obtained, it is possible to control IndSVs $Var(\hat{\beta}_k)$ via (2.6). Since $N \cdot IndPSV_k / \|x_k^*\|^2 = IndSV_k$ and since $q_k^2$ and $\check{R}_k^2$ do not depend on $\|x_k^*\|$, we can control $\|x_k^*\|$ by setting its lower bound for models in the class $D^{cd}$ as $D^{cde} = D^{cd} \cap D_3^e$ with

$$D_3^e = \{X^* \in [X] \mid \|p_k X^*\| \geq e, \ (k = 2, \cdots, J(X^*))\}.$$

Since any model in the set $D^{cd}$ satisfies

(3.4)  $H(I_k, VIF_k) \equiv I_k (p_k X^*) C_k(X^*) \leq cd \equiv \Delta_{cd}$

for any $k$, when model selection is restricted to the class $D^{cde}$, the SE in (2.7a), which is traditionally used as a measure of the stability of $\hat{\beta}_k$, is also controlled as

$$SE\left(\hat{\beta}_k(X^*)\right) = \hat{\sigma}(y^o, X^*) L(I_k, C_k)^{1/2} / \sqrt{N} \quad \text{with}$$

$$L(I_k, C_k) \equiv H(I_k, C_k) / \|x_k^*\|^2 \leq cd/e,$$

unless $D^{cde} = \emptyset$. Hence the upper bound for the SE is less than $\hat{\sigma}_{\hat{u}^*}\sqrt{\Delta_{cd}}/\sqrt{eN}$ for any $k \geq 2$, though the SEs are not comparable. While, since the IndPSVs is bounded above by $\sigma^2 \Delta_{cd}$ from (3.4), all the PSEs in (2.7b) are below or equal to $\hat{\sigma}_{\hat{u}^*}\sqrt{\Delta_{cd}}/\sqrt{N}$ where $\hat{\sigma}_{\hat{u}^*}$ is the RSE $\hat{\sigma}(\boldsymbol{y}^o, \boldsymbol{X}^*)$.

3.3 Finding $D^{cd}$ and Inefficiency-Collinearity Risk Index (ICRI) of $\boldsymbol{X}^*$

Now let us consider some methods of finding the class $D^{cd}$. One possible method is to try and check each of all the possible models, though its computational process may not be feasible if $K$ is large. But if an algorithm is available for finding *C*-controlled class $D_2^d$ in (3.3), the computational burden will be greatly reduced.

Empirically it is often the case that the set of variables in $\boldsymbol{X}$ is divided into core part $\boldsymbol{X}_1$ and such non-core part $\boldsymbol{X}_2$ as dummy variables as in Fox and Monette (1992). In such a case, it may be possible that a set of *IC*-controlled models is first found from the core part $\boldsymbol{X}_1$ and then each remaining variable in $\boldsymbol{X}_2$ is added to the set one by one.

Here, to take a different approach, we define the ICRI in order to compare different models in view of the EEM-M and apply it to finding *IC*-controlled models. In the case of the largest model $\boldsymbol{X}: N \times K$, the ICRI is defined as

(3.5a) $\quad r(\boldsymbol{X}) = (c_{MXK}, d_{MXK})$ with

(3.5b) $\quad c_{MXK} = \max_{k \geq 2} I_k(p_k \boldsymbol{X})$ and $d_{MXK} = \max_{k \geq 2} C_k(\boldsymbol{X})$.

Then it follows from (3.4) that

$$\max_k H(I_k(p_k \boldsymbol{X}), C_k(\boldsymbol{X})) \leq c_{MXK} d_{MXK}.$$

The right side of this inequality shows the worst performance of $\boldsymbol{X}$ in the criterion of the IndPSV. Clearly, if it is close to 1, the full model itself is *IC*-controlled and so is any sub-model without finding $D^{cd}$ because $D^{cd} = [\boldsymbol{X}]$. Hence, if $D^{cd} \neq [\boldsymbol{X}]$, the prespecified thresholds $c$ and $d$ in (3.3) need to be respectively set less than $c_{MXK}$ and $d_{MXK}$ and then we can apply one of the algorithms in Section 4 for seeking a practically useful class.

The risk indexes $c_{MXK}$ and $d_{MXK}$ in (3.5) are easily computed. In fact, $c_{MXK}$ is easily obtained from $K - 1$ individual variables $\{\boldsymbol{x}_k\}$ and $d_{MXK}$ is obtained by running $K - 1$ regression analyses among the $K - 1$ predictors to get $\check{R}_k^2(\boldsymbol{X})$'s $(k = 2, \cdots, K)$. Hence, the distribution of $(I_k(p_k \boldsymbol{X}), C_k(\boldsymbol{X}))$ for $k = 2, \cdots, K$ can be easily observed, through which an insight into the structure of collinearity will be obtained for selecting $(c, d)$ with expertized knowledge on the variables in $\boldsymbol{X}$. But it is easier to treat $(q_k^2, \check{R}_k^2)$ because these are squared correlations. Hence, to first select $(c_q, d_R)$, it is important to investigate the graph of $G \equiv \{(q_k^2(p_k \boldsymbol{X}), \check{R}_k^2(\boldsymbol{X})) | k = 2, \cdots, K\}$ and observe an empirical tendency on the relational structure of the whole predictors, from which we may be able to select them

judgmentally. Probably, both of $c_q$ and $d_R$ will be selected to be not less than 0.9.

First, to consider the problem of selecting $c_q$, we compute $\{q_k^2 | k = 2, \cdots, K\}$ once for all easily by using Proposition 2.2 (1). And then, since the condition $I_k > c$ is equivalent to the condition on the squared raw correlation:

(3.6) $\qquad q_k^2 \equiv (x_k'e)^2/[x_k'x_k \cdot e'e] > c_q \quad$ with $\quad c_q = 1 - 1/c$,

setting, e.g., $c_q = 0.95$, those $x_k$'s satisfying (3.6) may be deleted completely from the set of whole variables in $X$ in the first step. If $c_q = 0.9$ (or $0.95$), $c = 1/(1 - c_q) = 10$ (or 20 respectively). This will be one possible choice of $c$ for $I_k$. Then, without loss of generality, the remaining variables are renumbered and are denoted by $X(1) = \{x_1, x_2, \cdots, x_{K(1)}\}$ where $c_{MX(1)K(1)} = \max_k I_k(p_k X(1)) \leq c$ holds. Let $[X(1)] = \{X_\tau^{(1)}\}$ denote the set of all the subsets of $X(1)$, which are formed as matrices with $x_1 = e \in X_\tau^{(1)}$. Then all the models in $[X(1)]$ are $I$-controlled with $c_{MX(1)K(1)} \leq c$.

Second, once the inefficiency-risk is controlled, the rest part will go through with an algorithm for finding a class of $IC$-controlled models with $C_k \leq d$ for all $k$'s from $[X(1)]$. In Section 4, two algorithms are proposed. In this collinearity-controlling part, it will be easier to select $d_R = 1 - 1/d$ in terms of the CD (coefficient of determination) with $\check{R}_k^2(X) \leq d_R$ and the choice of $d_R$ can be made by such numbers as 0.9 or 0.95 from its meaning.

In general, studying the density of $\check{R}_k^2(X)$'s in a neighborhood of $\max_k \check{R}_k^2(X) = 1 - (1/d_{MXK})$ will be important not only in selecting $d$ but also in understanding the collinear structure of $X$.

3.4 ICRI and Admissible $y^o$-Accommodating Model $X$

In the same way as in (3.5), the ICRI of sub-model $X^*$ can be defined. But in treating sub-models, the column size effect on each $C$-index needs to be taken into account. First, recall that $[X; J]$ is the set of $N \times J$ models $\{X^*\}$ with $rank(X^*) = J$, $x_1 = e \in X^*$ as in (3.1). And let $X_{\tau J}^* \in [X:J]$, where $J = 2, \cdots, K$. Then, the ICRI of $X_{\tau J}^*$ is defined as

(3.7a) $\qquad r(X_{\tau J}^*) = (c_{M\tau J}, d_{M\tau J}) \quad$ with

(3.7b) $\qquad c_{M\tau J} = \max_k I_k(p_k X_{\tau J}^*) \quad$ and $\quad d_{M\tau J} = \max_k C_k(X_{\tau J}^*)$.

The ICRI $(c_{M\tau J}, d_{M\tau J})$ of model $X_{\tau J}^*$ is the "minimum" control level of $(c, d)$ for which $X_{\tau J}^*$ can belong to $D^{cd} \equiv D^{cd}([X])$ and hence if $c$ (or $d$) in (3.3) is smaller than $c_{M\tau J}$ (or $d_{M\tau J}$ resp.), then $X_{\tau J}^*$ is excluded from $D^{cd}$. It is noted that $\check{R}_k^2(X_{\tau J}^*) \leq 1 - d_{M\tau J}^{-1}$. Let the model with the ICRI in (3.6a) be denoted as $X_{\tau J}^*(c_{M\tau J}, d_{M\tau J})$. Since $d_{M\tau J}$ depends on the column size $J$ as well as matrix $\tau$, the set $D^{cd}$ in (3.3) is disjointly decomposed by the column size as

(3.8a) $D^{cd} = \cup_{J=2}^{K*} D_J^{cd}$ with

(3.8b) $D_J^{cd} \equiv \{X_\tau^* \in [X:J] | I_k(p_k X_\tau^*) \leq c, C_k(X_\tau^*) \leq d, (k = 2, 3, \cdots, J)\}$.

Then, if $X_{\tau J}^* \in D_J^{cd}$ is given,

(3.9a) $\quad IndPSV_{\tau k J} = \sigma^2 I_{\tau k J} C_{\tau k J}/N \leq \sigma^2 \min\{H_{M\tau J}, cd\}/N$

for any $k$, where $H_{M\tau J} = c_{M\tau J} d_{M\tau J}$. Hence averaging this over $k$'s yields the upper bound for the averaged IndPSV of the model:

(3.9b) $\quad \sigma^2 \overline{H_{\tau J}}/N = \overline{IndPSV_{\tau J}} \leq \sigma^2 \min\{H_{M\tau J}, cd\}/N$,

showing the upper bound for the mean control level of IndPSV for model $X_{\tau J}^*$.

Next, the concept of admissibility is defined on the set $[X:J]$ with rank $J \geq 2$.

Definition 3.2  For each $J$ fixed, $X_\tau^*(c_{M\tau J}, d_{M\tau J})$ is said to better accommodate $y^o$ than $X_{\tau\prime}^*(c'_{M\tau J}, d'_{M\tau J})$ if $c_{M\tau J} \leq c'_{M\tau J}$ and $d_{M\tau J} \leq d'_{M\tau J}$ hold with one of the inequalities strict. If no matrix of the same column size accommodates $y^o$ better than $X_\tau^*(c_{M\tau J}, d_{M\tau J})$, then $X_\tau^*(c_{M\tau J}, d_{M\tau J})$ is said to be admissible in $D_J^{cd} \equiv D_J^{cd}([X])$. Hence, the minimal accommodation set for $y^o$ is defined by the set of admissible matrices in $D_J^{cd}$;

$$D_{admJ}^{cd} \equiv D_{admJ}^{cd}([X]) = \{X_\tau^* \in D_J^{cd} | X_\tau^* \text{ is admissible}\}.$$

Any model in $D_{admJ}^{cd}$ is called an admissible $y^o$-accommodating model in $[X:J]$.

For $J$ fixed, it will be desirable to have such a minimal set as $D_{admJ}^{cd}$ for a possible model selection and all the admissible models are those in the set $D_{adm}^{cd} \equiv \cup_{J=2}^{K*} \{D_{admJ}^{cd}\}$, which forms a complete class (see Lehmann and Romano (2005)). However, it is not necessarily required to have a model in this set. In fact, as IndPSVs are only required to be uniformly controlled, any model in the $y^o$-accommodating set (3.3) will do, and it will be important to select a model for accommodating $y^o$ in an effective and balanced way so that *I*-index and *C*-index are acceptably well controlled. Furthermore, as was discussed in 3.1, a final model should be selected by a $y^o$MSP together with such criteria as AIC, BIC, RSE, CD that use $y^o$. The XMSP simply gives a set of models whose two indexes are uniformly controlled.

Example 3.1  Let $J=2$. Then the CD where $x_k$ is regressed on $x_1 = e$ is regarded as the normalized squared inner product $r_{1k}^2 = (x_k'e)^2/[x_k'x_k \cdot e'e] = \cos^2(\theta_k)$ as the collinearity index with $C_k = (1 - r_{1k}^2)^{-1}$. By Proposition 2.2, $I_{k2}$ and $C_{k2}$ are interdependent, and $H_{k2} = (1 - r_{1k}^2)^{-2}$. Here $k$ denotes the $k$-th matrix $X_k^* = (x_1, x_k)$ of $N \times 2$.

Proposition 3.2  Let $J=2$ and let $\boldsymbol{X}_k^* = (\boldsymbol{x}_1, \boldsymbol{x}_k) \in [\boldsymbol{X}\!:\!2]$ with $k = 2, 3, \cdots, K$. If $\boldsymbol{X}_i^* = (\boldsymbol{x}_1, \boldsymbol{x}_i) \in D_2^{cd}$ satisfies $H_{i2} = (1 - r_{1i}^2)^{-2} = \min_k (1 - r_{1k}^2)^{-2}$, it is admissible, where $r_{1i}^2 = \min_k r_{1k}^2$.

In fact, $\boldsymbol{X}_i^* = (\boldsymbol{x}_1, \boldsymbol{x}_i) \in D_2^{cd}$ satisfies $I_{i2} \cdot C_{i2} = \min_k I_k \cdot C_{k2} = \min_k (1 - r_{1k}^2)^{-2}$, because no other model is better than this model.

This is not an example for controlling the two indexes separately. As in Proposition 2.2(2), in these simple models $I_k$ and $C_k$ are not separable and they are simultaneously controlled by minimizing $H_{k2} = (1 - r_{1k}^2)^{-2}$. Hence, it will be better to have a model with smaller $r_{1i}^2$, which makes the two indexes simultaneously better controlled. However, in any regression model, it may be better to select one from $D_2^{cd}$ by using $y^o$MSP together with a judicious judgement.

4  Collinearity-Controlling Algorithms for XMSP

In this section, we materialize the XMSP by developing two algorithms in order to derive accommodating classes for $y^o$MSP with focus on controlling collinearity. As has been stated in Section 3.2, to get a $D^{c,d} = D_1^c \cap D_2^d$ of IC-controlled models for given control parameters $(c, d)$, the I-controlled class $D_1^c$ in (3.3a) is first and easily obtained via Proposition 3.2 because $I_k \leq c$ is equivalent to $q_k^2 \leq c_q = 1 - 1/c$. Then $D_1^c$ is the class of all the models $\{\boldsymbol{X}_\tau\}$ such that each variable $\boldsymbol{x}_k$ in $\boldsymbol{X}_\tau$ satisfies $q_k^2 \leq c_q$, where $k \geq 2$, and $\boldsymbol{e} \in \boldsymbol{X}_\tau$. Here $q_k^2$ is the squared raw correlation of $\boldsymbol{x}_1 = \boldsymbol{e}$ and $\boldsymbol{x}_k$ (see (3.6)). Then $\overline{\boldsymbol{X}} \equiv \cup_\tau \boldsymbol{X}_\tau$ also belongs to $D_1^c$ and so it suffices to find the class $D_2^d$ of C-controlled models in $\overline{\boldsymbol{X}}$ that belong to $D_2^d$. Hence, without loss of generality, regarding $\boldsymbol{X}$ as $\overline{\boldsymbol{X}}$, the algorithm is considered in terms of the coefficient of determination $CD_k \equiv \check{R}_k^2$, which is the $CD_k$ when $\boldsymbol{x}_k$ is regressed on $\check{\boldsymbol{X}}_k = \{\boldsymbol{x}_j \in \boldsymbol{X}, j \neq k\}$.

The two algorithms for obtaining $D_2^d$ are described below: variable-increasing algorithm (VI-Algorithm) in Section 4.1 and variable-reducing algorithm (VR-Algorithm) in Section 4.2. In the latter case, principal component analysis (PCA) will be used, where PCA here is nothing but an orthogonal diagonalization of the correlation matrix of $\boldsymbol{x}_j$'s in order to select groups of variables that are strongly correlated with the leading principal components. In the analysis, it is shown that $\check{R}_k^2$ based on $\boldsymbol{X}$ is equal to $\check{R}_k^2$ based on the $N \times (K - 1)$ matrix $\boldsymbol{Z}^*$ of standardized variates of $\boldsymbol{x}_j$ ($k \geq 2$). Note that by Proposition 2.1, if $CD_k(\boldsymbol{X}^*) \leq d_R$ for any $k$, then $CD_k(\boldsymbol{X}^{**}) \leq d_R$ for any sub-model $\boldsymbol{X}^{**}$ of $\boldsymbol{X}^*$. In Section 4.2, some analytical results are made on the upper and lower bounds for pairwise correlations in each principal component collinearity (PCC) class of variables. In Section 5, a numerical example of the XMSPs that uses the two algorithms is given.

4.1 Variable-increasing Algorithm (VI-Algorithm).

A VI-Algorithm is here developed for controlling strong collinearity in $X$ in advance to the traditional OLS analysis via $y^o$MSP. Let $J = \{2,3,\cdots,K\}$, and let

(4.1)  $G(p: A_{p-1}) = \{(i_h^p, A_{p-1}): CD(i_h^p, A_{p-1}) \le d_R, \ i_h^p \in J\backslash A_{p-1}\}$

denote the set of suffixes $i_h^p{}'s$ of $x$ variables in the $p$-th step after the set $A_{p-1}$ of $x$'s suffixes has been selected in the ($p$-1)-th step, where $CD(i_h^p, A_{p-1}) \le d_R$ denotes the CD condition for the variable of suffix $i_h^p$ to satisfy. Here $CD(i_h^p, A_{p-1})$ is the $CD$ -value when $x_{i_h^p}$ is regressed on the set $X(A_{p-1})$ of the $x$ variables with suffixes in $A_{p-1}$. The set $J_p$ of available suffices, the set $A_p$ of selected suffices and the whole picture in each step respectively move as follows.

$J_{p+1} = J_p \setminus G(p+1: A_p)$ with $J_1 = \{2,\cdots,K\}$ and $A_1 = \{1\}$.

$A_p \equiv A_p(i_h^p) = \{i_h^p\} \cup A_{p-1}$ for $i_h^p \in G(p: A_{p-1}), p \ge 2$, and

$\begin{Bmatrix} A_1 \\ J_1 \end{Bmatrix} \to G(2: A_1) \to \begin{Bmatrix} A_2 \\ J_2 \end{Bmatrix} \to G(3: A_2) \to \begin{Bmatrix} A_3 \\ J_3 \end{Bmatrix} \to G(4: A_3) \to \cdots$ .

Here $A_1 \subset A_2 \subset A_3 \subset \cdots$, $J_1 \supset J_2 \supset J_3 \supset \cdots$, and $G(p: A_{p-1}) \subset A_p$.

It is noted that newly added $x$ variables through the selected suffixes in the $p$-th step have $CD$'s with the variables $X(A_{p-1})$ that are smaller than or equal to $d_R$, securing stability.

Using this notation, we describe our VI-Algorithm.

(Step1) Let $A_1 = \{1\}$ in the first step, implying that $x_1$ is always included, which corresponds to the first step. Let $J_1 = \{2,\cdots,K\}$.

(Step 2) Then the suffixes selected in the second step is given by

(4.2)  $G(2: A_1) = \{(i_h^2: A_1): CD(i_h^2, A_1) \le d_R, \ i_h^2 \in J_1\backslash A_1\}$.

Here the $CD$ of $x_k$ with $\{x_1\}$ is defined by the square raw correlation

(4.2a)  $CD_k(\{x_1\}) = [x_k'x_1/\|x_k\|\|x_1\|]^2 \ (\equiv q_k^2),$ ,

which is used only for this second step. Let $G(2: A_1)$ be extensively expressed as

(4.3)  $G(2: A_1) = \{(i_h^2, \{1\}) | i_h^2 = i_1^2, \cdots, i_{l(2)}^2, 1 < i_1^2 < \cdots < i_{l(2)}^2\}$.

Then each of the pairs $\{(x_i, x_1): i = i_1^2, \cdots, i_{l(2)}^2\}$ satisfies the $CD$ condition and hence each pair defines $A_2(i_k^2) = \{i_k^2, 1\}$ ($k = 1, \cdots, l(2)$). Note $l(2) \le K - 1$. Let $J_2$ be the suffix set of the $(K - 1 - l(2))$ remaining variables, since in this step $l(2)$ suffixes (and so variables) are deleted from $J_1$.

Example 4.1 If the suffixes $i_1^2 = 3, i_2^2 = 5, i_3^2 = 6, i_4^2 = 7, i_5^2 = 9$ and $i_6^2 = 10$ satisfy the CD condition in $J_1 = \{2, \cdots, K\}$ and the others do not, then the suffix set $J_2 = \{3,5,6,7,9,10\}$ is carried over to the next choice set, while the suffixes in the set $\{2,4,8\}$ are deleted completely. Here $l(2) = 6$.

In step 2, variables with the suffixes in the complement set $G(2:A_1)^c$ do not satisfy the condition and so those variables are deleted from the model completely so long as $x_1 = e$ is included as a part of the model.

(Step 3) The third step concerns the $i_{l(k)}^2$ suffix sets for $k = 1, \cdots, i_{l(k)}^2$ given by

(4.4) $\quad G\left(3: A_2(i_k^2)\right) = \{(i_h^3, A_2(i_k^2)): CD(i_h^3, A_2(i_k^2)) \le d_R, \ i_h^3 \in J_2 \backslash A_2(i_k^2)\}$

$\quad \equiv \{(i_h^3, A_2(i_k^2)) \mid i_h^3 = i_1^3, \cdots, i_{l(3)}^3, 1 < i_1^3 < \cdots < i_{l(3)}^3\},$

where $l(3) \le K - 1 - l(2)$. In other words, for each choice of $A_2(i_k^2) = \{i_k^2, 1\}$ ($k = 1, \cdots, l(2)$), the $A_3(i_j^3: A_2(i_k^2))$ is defined as the three suffixes.

(4.5) $\quad A_3(i_j^3: A_2(i_k^2)) = \{i_j^3: A_2(i_k^2)\} = \{i_j^3, i_k^2, 1\}$

For easy exposition, let us stay in Example 4.1 and suppose $i_2^2 = 3$. Then such $\{i_j^3 = j\}$ is chosen so that $CD(j, \{3,1\}) \le d_R$ when $x_j$ regressed on $\{x_3, x_1\}$ for $j \in \{5,6,7,9,10\}$. While, if $i_2^2 = 5$, such $\{i_j^3 = j\}$ is chosen so that $CD(j, \{5,1\}) \ge d_R$ when $x_j$ regressed on $\{x_5, x_1\}$ for $j \in \{3,6,7,9,10\}$. Note that $CD(5, \{3,1\})$ and $CD(3, \{5,1\})$ are different.

(Step 4) For each $i_k^2 \in J_2 = \{3,5,6,7,9,10\}$, there are the 6 suffix sets $\{i_j^3: A_2(i_k^2)\}$ with $i_j^3$ satisfying $CD(i_h^3, A_2(i_k^2)) \le d_R$. Then for each of these 6 sets as in (5.6), we find

(4.6) $\quad G\left(4: A_3(i_j^3)\right) = \{(i_h^4: A_3(i_j^3)): CD\left(i_h^4, A_3(i_j^3)\right) \le d_R, \ i_h^4 \in J_3 \backslash A_3(i_j^3)\}.$

This branching process stops when no suffix satisfying the CD condition is found. And then we will have the set of homogeneous suffix subsets $\{A_d^*\}$ each of which corresponds to a subset $X_d^*$ of variables satisfying the CD condition within $X_d^*$.

We remark that in the case of variable-increasing method, it is not necessary to first find the class $D_1^c$ of variables that satisfy $q_k^2 \le c_q$ so long as $c_q$ equals $d_R$ in $CD_k \le d_R$ because checking $q_k^2 \le c_q$ is included in step 1.

4.2 Variable-reducing Algorithm (VR-Algorithm) via PCA.

As an alternative method for selecting an empirically effective model, we propose a VR-Algorithm based on PCA on data $X$. First, via PC method, we will make what we call principal component collinearity (PCC) classes of variables (e.g., the *m*-th PCC class $G_{ma}$), such that each variable in $G_{ma}$ has a strong correlation in absolute value with the *m*-th PC. The PCC classes $\{G_{ma}\}$ are mutually disjoint and variables in each $G_{ma}$ linearly and strongly cluster along the corresponding PC. In each class, the correlation in absolute value between each variable and the PC is controlled to be more than a prespecified level $a$ ($\geq 0.9$), so that the squared correlation of any two variables in the class $G_{ma}$ will be bigger than or equal to a certain level and the squared correlation of any two variables taken from two different PCC classes becomes close to zero. Therefore, in view of the EEM, two or more variables in each class should not be used simultaneously in a same regression model to avoid a strong collinearity, so long as the class contains more than one variable.

(1) To develop the procedure, first, we show that for fixed $k$, the CD $\check{R}_k^2 \equiv \check{R}_k^2(X)$ ($k \geq 2$) is expressed in terms of standardized variables $\{z_j\}$, where

(4.7a) $\quad z_j = (z_{nj}), \quad z_{nj} = (x_{nj} - \bar{x}_j)/\sqrt{N}s_{xj} \quad (j \geq 2)$ and

(4.7b) $\quad Z^* = (z_2, \cdots, z_K): N \times (K-1)$.

Let $\check{R}_k^2(Z^*)$ be the CD of $z_k$ that is regressed on $\check{Z}_k^*$ ($k \geq 2$). Then the following lemma holds, where the proof is given in Appendix 1.

Lemma 4.1 $\check{R}_k^2(X) = \check{R}_k^2(Z^*)$ ($k \geq 2$).

An important implication of this lemma is that the collinearity of variables $\{x_k\}$ in $X$ completely corresponds to that of standardized variables $\{z_k\}$ in $Z^*$ where $x_1 = e$ and $e \in \check{X}_k$ for any $k \geq 2$.

(2) Next, analyzing $Z^*$ via PC method, we will find disjoint PCC classes of variables such that variables in each class cluster linearly and strongly along the corresponding PC.

To describe the procedure, let $Z^{*\prime}Z^* = P\Lambda P'$ with $\Lambda = diag\{\lambda_2, \cdots, \lambda_K\}$ be the orthogonal decomposition of $Z^{*\prime}Z^*$, where $\lambda_2 \geq \cdots \geq \lambda_K$ are the latent roots of $Z^{*\prime}Z^*$ with $\lambda_2 + \cdots + \lambda_K = K - 1$. And $P = [p_2, \cdots, p_K]$ is a (*K*-1)×(*K*-1) orthogonal matrix with $p_j$ the *j*-th latent vector corresponding to $\lambda_j$. If some of $\lambda_j's$ are equal, $P$ has different expressions in addition to sigh changes of $p_j$'s. In such a case, we pick one and fix it throughout the analysis. Furthermore, we need some more notations.

First, setting $F = Z^*P$ and $p_j = (p_{2j}, \cdots, p_{Kj})'$ yields $Z^* = FP'$ and $F'F = \Lambda$. Let $F = [f_2, \cdots, f_K]$. Then, $f_k'f_j = 0$ ($k \neq j$) and $f_k'f_k = \lambda_k$ follow, and variables are expressed as $f_j = z_2 p_{2j} + \cdots + z_K p_{Kj}$ and

(4.8) $\quad z_k = f_2 p_{k2} + f_3 p_{k3} + \cdots + f_K p_{kK}$

$\quad\quad\quad = g_2 \sqrt{\lambda_2} p_{k2} + g_3 \sqrt{\lambda_3} p_{k3} + \cdots + g_K \sqrt{\lambda_K} p_{kK},$

where $g_j = f_j / \sqrt{\lambda_j}$ is the standardized component with mean 0 and variance 1 as $e' g_j = 0$ and $g_j' g_j = 1$. In this formulation, $f_j$ should be the ($j$-1)-th principal component of $Z^{*'} Z^*$, but having understood it, we shall call $f_j$ the $j$-th component for $j \geq 2$ by keeping its original name of the suffix. Then, since $e' z_k = 0$ and $z_k' z_k = 1$,

(4.9) $\quad g_j' z_k / \|g_j\| \|z_k\| = \sqrt{\lambda_j} p_{kj}$

is the correlation of the $k$-th variable $z_k$ and the $j$-th standardized PC $g_j$.

Next, let

(4.10) $\quad \delta_j = \lambda_j / (K - 1) \quad (j \geq 2)$

be the relative contribution of the $j$-th PC to the total variation $(K - 1)$. In the sequel, we focus on the 2nd PCC class of variables because $\delta_2 \geq \delta_3 \geq \cdots \geq \delta_K$. The larger $\delta_2$ is, the more strongly the 2nd component $g_2$ impacts on the total variations of variables $\{z_k\}$. This implies that when $\delta_2$ is large, those variables $\{z_k\}$, which have larger absolute correlations $|\sqrt{\lambda_2} p_{k2}|$'s with $g_2$, tend to be more strongly correlated within the variables. In fact, from (4.8) the correlation of $z_k$ and $z_j$ is decomposed as

(4.11a) $\quad z_k' z_j = \lambda_2 p_{k2} p_{j2} + \lambda_3 p_{k3} p_{j3} + \cdots + \lambda_K p_{kK} p_{jK},$

(4.11b) $\quad 1 = z_k' z_k = \lambda_2 p_{k2}^2 + \lambda_3 p_{k3}^2 + \cdots + \lambda_K p_{kK}^2$

and so if $|\sqrt{\lambda_2} p_{i2}|$ is large for $i = k, j$, the correlation $z_k' z_j$ of $z_k$ and $z_j$ is large in absolute value due to the common component $g_2$ and they are less correlated with those variables that have small correlation with $g_2$.

Now, we describe our approach to controlling strong collinearity in the VR-Algorithm. First using (4.9), we define the statistic of the $m$-th collinearity identifier of the $k$-th variable $z_k$ by

(4.12) $\quad d(m, k) \equiv d(m, k : Z^{*'} Z^*) = |\sqrt{\lambda_m} p_{km}| \quad (m, k = 2, \cdots, K)$

which is the absolute value of the correlation between $z_k$ and $g_m$, and define the $m$-th principal component collinearity (PCC) class $G_{maz}$ of variables at level $a$ by the set of variables given by

(4.13a) $\quad G_{maz} = \{z_k \in Z^* | d(m, k) \geq a\} \quad (a \in [0.9, 1]),$

where the threshold variable $a$ for the $m$-th collinearity identifier is prespecified in advance and assumed to belong to the interval [0.9, 1] so that a strong collinearity may be avoided in regression matrix $X$ as will been shown below. Also, let $\#(G_{maz}) \equiv h_m$ denote the number of variables contained in $G_{maz}$. Then, it is easy to observe that

1) for given $a \in [0.9, 1]$, the larger the variance $\lambda_m$ of the $m$-th PC is, the more strongly the variables in $G_{maz}$ cluster together along the $m$-th PC, or equivalently the more strongly they are collinear, so long as $\#(G_{maz}) \equiv h_m \geq 2$ and
2) the closer the threshold $a$ is to 1, the smaller the $m$-th PCC class is, but the more strongly they are collinear in terms of correlations among variables in the class, so long as $\#(G_{maz}) \equiv h_m \geq 2$.

By the definitions in (4.12) and (4.13a), the fact 2) implies that $z_k$ belongs to $G_{maz}$ for any $|p_{km}|$ (or equivalently for any $Z^*$) if and only if $\lambda_m \geq 1$. Hence, $\lambda_m \geq 1$ is a necessary condition for the $m$-th PCC class to be treated as a collinearity class of variables. In addition, by the fact 1), the variables in the 2$^{nd}$ PCC class $G_{2az}$ with $\lambda_2$ cluster linearly along the 2$^{nd}$ PC, implying a strong collinearity of the variables, though it is not directly measured in the CDs. In fact, by (4.9) and (4.12), each $z_k$ in

(4.13b) $\quad G_{2az} = \{z_k \in Z^* | \, d(2,k) \geq a\}$,

has CD $d(2,k)^2$ in the linear relation between $z_k$ and $g_2$. If $a = 0.95$, $d(2,k)^2 \geq 0.903$. For the correlation between $z_k$ and $z_j$ in $G_{2az}$, the following lemma holds from (4.8) and (4.11a). The proof is given in Appendix.

**Lemma 4.2** Suppose that $\#(G_{maz}) \equiv h_m \geq 2$. For $z_k$ and $z_j$ ($k \neq j$) in $G_{maz}$, the correlation $|z'_k z_j|$ in absolute value satisfies

(4.14a) $\quad L_{mkj} \leq |z'_k z_j| \leq d(m,k)d(m,j) + (1 - a^2)$ with

(4.14b) $\quad L_{mkj} \equiv d(m,k)d(m,j) - (1 - a^2)$,

where $F(a) \equiv 2a^2 - 1 \leq L_{mkj}$.

The lower bound $L_{mkj}$ for $|z'_k z_j|$ depends on the two absolute correlations of $d(m,k)$ and $d(m,k)$. In a real data example of Section 5.2, it will be shown that $d(2,2) = 0.980$ and $d(2,3) = 0.957$ and so $L_{2.2.3} = 0.826$ for $|z'_2 z_3|$. While, the lower bound $F(a)$ for $L_{mkj}$ in (4.14b) is common to all the PCC classes and so in any PCC class, when $a = 0.95$, $F(a) = 0.805$ if the class is not empty. Conversely, when $a = 0.95$, no two variables whose absolute correlation is less than 0.805 cannot be the variables in a same PCC class.

In the VR-Algorithm with $a \in [0.9, 1]$, it is required that the variables in $G_{2az}$ be not used simultaneously in a same model if $\#(G_{2az}) \equiv h_2 \geq 2$.

It is noted that since $a \geq 0.9$, $G_{jaz} \cap G_{maz} = \emptyset$ for $j \neq m$ because of (4.11b) and $p_{km}$ is the $k$-th element of $\boldsymbol{p}_m$. As will be shown below, there will not be many PCC classes such that $h_m \geq 2$ for $a \geq 0.9$.

(3) The VR-Algorithm for getting empirically effective models via PCA.

Now we develop an MSP in EEM. Here, we will add another condition on the sizes of $\delta_j's$ to make the collinearity of the variables in each PCC class stronger. The procedure for getting a set $\{G_{maz}\}$ of PCC classes will be stopped at $m = M$, where

(4.15a)     $M$ is the least value for which $\delta_M \geq b$.

In fact, the choice of $M$ in (4.15a) requires that the relative contribution of the $j$-th PC is at least larger than or equal to $b$, where $\delta_2 \geq \delta_3 \geq \cdots \geq \delta_K$ with $\delta_j's$ in (4.10). Note if we set $b = 0.4$, $M$ is at most 2. This implies that there are at most 2 PCC classes of variables such that variables in each class is likely to cause strong collinearity in $\boldsymbol{Z}^*$. It is remarked that if $\delta_M \geq 0.4$, $\lambda_M = (K-1)\delta_M > 1$ is necessary for $K \geq 4$, and if $\delta_M \geq 0.4$ and $K - 1 = 10$, $\lambda_M \geq 4$ is necessary.

Here, to avoid a complex algorithm, we simply describe it only for the case of

(4.15b)     $a = 0.9$ and $\delta_2 \geq \delta_3 \geq b = 0.4$ or equivalently $M = 2$.

A general case can be easily extended in the same manner and hence the details of the procedure are omitted. By this setting, the first two PCC classes are related to the VR-Algorithm to control strong collinearity.

If both the number $h_2$ and $h_3$ of variables contained in $G_{2az}$ and $G_{3az}$ are less than or equal to 1, nothing has to be done concerning strong collinearity, so long as the variables are $I$-controlled. In such a case, there is only one model $\boldsymbol{X}_\tau \equiv \boldsymbol{X}$ and then the procedure goes to the VR-process in (4).

If $h_2 \geq 2$ and $h_3 \leq 1$ or if $h_2 \leq 1$ and $h_3 \geq 2$, then these cases are treated in a similar manner as in the next case. Assume $h_2 = 3$ and $h_3 = 2$, the correlations of variables in $G_{2az}$ and $G_{3az}$ with the 2nd PC and the 3rd PC are respectively more than or equal to 0.9. Then, no two variables in each PCC class cannot be included in one model and so there are $h_2 \times h_3 = 6$ starting groups of variables to form model candidates in our VR-Algorithm.

So far, the PCA was discussed for selection of variables in terms of $\boldsymbol{Z}^*$, but by Lemma 4.1, all the arguments go through with $\boldsymbol{x}_k$ for $\boldsymbol{z}_k$ so long as the problem of controlling CDs is concerned. Let $Q_j$ ($j = 1, \cdots, 6$) be the $j$-th group selecting one variable each from $G_{2az}$ and $G_{3az}$ such that $Q_j \cap Q_k = \emptyset$ ($j \neq k$). To select model candidates on which the collinearity condition CD $\leq d_R$ is imposed, let $H_M = X - \cup_{j=1}^{6} Q_j$ and define 6 model candidates by

$$X_\tau = Q_\tau \cup H_M \quad (\tau = 1, \cdots, 6).$$

A detailed process is given below for getting an empirically effective model from each $X_\tau$ via VR-Algorithm in terms of CDs, where for simplicity, each $X_\tau$ is regarded as $X: N \times K$, though in fact, $K$ here corresponds to $K - 6$.

(4) Controlling CDs in each $X_\tau$ to select a class of models with $\max_k CD_k \leq d_R$.
(Step 1) For a given $X(0) = X$, let $\Xi_0^{(0)} = \{CD_k^{(0)}: k = 2,3, \cdots, K\}$, $x_k^{(0)} = x_k$, $\breve{X}_k^{(0)} = \breve{X}_k$, and $CD_k^{(0)} = CD_k$. Suppose that $CD_K^{(0)}$ of $x_K^{(0)} \neq x_1 = e$ is the largest CD in $\Xi_0^{(0)}$ and $CD_K^{(0)} = \max_i CD_i^{(0)} > d_R$, then delete $x_K$ and replace $X(0)$ by $\breve{X}_K^{(0)} = \breve{X}_K$.

(Step 2) Denote the set of remaining variables by $X(1) = \breve{X}_K^{(0)} = \{x_1, x_2, \cdots, x_{K-1}\}$ via renumbering and let $\Xi_0^{(1)} = \{CD_k^{(1)}: k = 2, \cdots, K-1\}$. Suppose that $CD_{K-1}^{(1)}$ of $x_{K-1}^{(1)} \neq x_1 = e$ is the largest CD in $\Xi_0^{(1)}$ and $CD_{K-1}^{(1)} = \max_i CD_i^{(1)} > d_R$. Then $X(2) = \breve{X}_{k-1}^{(1)} = \{x_1, x_2, \cdots, x_{K-2}\}$ is the set of $K$-2 variables after $x_{K-1}^{(1)}$ is deleted from $X(1) = \breve{X}_K^{(0)}$.

(Step 3) Repeating this variable-reducing process, we stop it at the $p$-th step when there is no CD such that $CD_i^{(p)} > d_R$. Then $X(p) = \breve{X}_{K-1+p}^{(p-1)} = \{x_1, x_2, \cdots, x_{K-p}\}$ by renumbering the suffixes is a final set for which the OLS analysis is applied without any serious effect of collinearity.

By Proposition 2.1, any sub-model from this set satisfies $\max_k CD_k \leq d_R$.
In practice, the condition $CD_K^{(0)} = \max_i CD_i^{(0)} \leq d_R$ required in the above process will be satisfied in the first step and also $h_3 = 0$ will hold in most data after the screening of variables via $q_k^2 \leq c_q$ has been carried out by the *I*-index.

In the case of VR-Algorithm, the parameters of $(c_q, d_R, a, b)$ is required to be prespecified. It is noted that $a$ is different from $d_R$ and a well setting of $(a, b)$ will play a role of seeking and controlling strong collinearity in confounding relations of variables. But the problem will be left out as a matter of analysts' choice, and it will need many empirical works to get a rule of thumb.

Thus, as the XMSP in our EEM-M, we have a class $D^{c,d}$ of empirically effective models and then one of the models will be selected from $D^{c,d}$ as a final model via $y^o MSP$ in empirical analysis in view of the EEM.

5 Numerical Examples of VI-Algorithm and VR-Algorithm
As a numerical example for demonstration, we here briefly describe the VI-Algorithm and VR-Algorithm in the gasoline consumption data (Y) in *Motor Trend magazine for the year 1975,* which is listed in Chatterjee and Hadi (2012) (Tables 9.16-9.17, pp255-256). The data is collected on 30 models of cars to study the factors that determine the gasoline consumption of cars. The variables are summarized in Table 5.1. Note *N*=30.

In addition, using the correlation structure of this data as given, a simulation analysis will be made for the VR-Algorithm as an experiment, where vector variables are assumed to be independently and identically distributed ($iid$) as normal distribution, though actual data is not generated under such assumptions.

Table 5.1 Variables for the Gasoline Consumption Data

| Y | Miles/gallon | X2 | Displacement (cubic inches) |
|---|---|---|---|
| X3 | Horsepower (feet/pound) | X4 | Torque (feet/pound) |
| X5 | Compression ratio | X6 | Rear axle ratio |
| X7 | Carburetor (barrels) | X8 | No. of transmission speeds |
| X9 | Overall length (inches) | X10 | Width (inches) |
| X11 | Weight (pounds) | X12 | Transmission type (1:automatic, 0:manual) |

(1) *I*-controlled class of variables

Let us first consider the effect of the *I*-index on the selection of variables in the XMSP. Since $I_k = 1/(1 - q_k^2)$ (see Proposition 2.2) holds for the squared raw correlation $q_k^2 \equiv CD_1(\{x_1\})$ (see (4.2a)), the condition $q_k^2 \leq c_q$ for inclusion of variables is equivalent to

$$I_k \leq 1/(1 - c_q),$$

where $I_k$ is the inefficiency index of the *k*-th estimate. As in Proposition 3.2, if the threshold is set to be $c_q = 0.9$ for $q_k^2$, then the threshold for $I_k$ becomes $c$ =10. Similarly, if $c_q = 0.95, then\ c = 20$

Table 5.2 The values of $I_k = 1/(1 - q_k^2)$ $(k = 2, \cdots, 12)$.

| $I_k$ | X2 | 7.14 | X3 | 11.1 | X4 | 8.33 | X5 | 1000 | X6 | 33.3 |
|---|---|---|---|---|---|---|---|---|---|---|
| $q_k^2$ | | 0.86 | | 0.91 | | 0.88 | | 0.999 | | 0.97 |
| X7 | 7.14 | X8 | 25 | X9 | 25 | X10 | 25 | X11 | 17.7 | X12 | 3.70 |
| | 0.86 | | 0.96 | | 0.99 | | 0.99 | | 0.94 | | 0.73 |

In Table 5.2, the values of $I_k$ and $q_k^2$ are computed. From this table, when $c_q = 0.9$, variables X3, X5, X6, X8, X9, X10 and X11 are deleted and the remaining variables are those of the set

(5.1)   X(1,0.9)={X2, X4, X7, X12}.

On the other hand, when $c_q = 0.95$, variables X5, X6, X8, X9 and X10 are deleted and the remaining variables are those of the set

(5.2)	$\mathcal{X}(1, 0.95) = \{X2, X3, X4, X7, X11, X12\}$.

As has been pointed out in Proposition 2.2 (3), the fact that the squared raw correlation $q_k^2$ of $\boldsymbol{x}_1 = \boldsymbol{e}$ and $\boldsymbol{x}_k$ is large means that the $k$-th variable $\boldsymbol{x}_k$ is more likely to be almost constant like $\alpha \boldsymbol{e}$ for some $\alpha$. For example, since $q_5^2$ is almost 1, implying $\boldsymbol{x}_5 \approx \alpha \boldsymbol{e}$ and that the compression ratio X5 does not much change over different cars. The inefficiency index turns out to be a measure of immobility and so it is not useful in regression with a constant term. Such variables will be deleted by screening them via the $I$-index. It is noted that the data of gasoline consumption is not used at all in the XMSP.

## 5.1 XMSP via VI-Algorithm

Now, when $c_q = 0.9$, the VI-Algorithm is applied to the model $\mathcal{X}(1, 0.9) = \{X2, X4, X7, X12\}$ in (5.1). As is described in 4.1, a new variable is added one by one to the set of the previously selected variables and if $\check{R}_k^2 > d_R$ in a step, the new variable is not included, otherwise it is added to the model that is explored. It starts with the pairs (X1, Xm) (m=2,4,7,12) where X1 is $\boldsymbol{x}_1 = \boldsymbol{e}$. The models we finally obtained under the current specification of parameters are the two models:

(5.3)	$\mathcal{M}(1) = \{X1, X2, X7, X12\}$ and $\mathcal{M}(2) = \{X1, X4, X7, X12\}$.

Here the variables X2 and X4 are competing variables for model candidacy in view of $\check{R}_k^2$. In fact, when X4 is regressed on $\mathcal{M}(1)$, the CD is 0.98, which is the reason why $\mathcal{M}(1)$ does not include X4 under $d_R = 0.9$. Thus, we obtain the $\boldsymbol{y}^o$-accommodating class of models:

(5.4)	$\mathcal{D}^{c_q=0.9, d_R=0.9} = \{\mathcal{M}(1), \mathcal{M}(2)\}$.

The larger each threshold of $c_q$ and $d_R$ is, the less stringent the variable selection becomes and the more candidate models probably with more variables in each model are obtained. Hence, it will be necessary to study more on this problem of choosing the threshold values $c_q$ and $d_R$. In general, it will be difficult to get this kind of the information under a compound collinear structure of variables.

Incidentally, in the case of $c_q = 0.9$ and $d_R = 0.9$, $H_k$ in (2.3) is bounded above by 10 ×10=100, which may be too big. Though, the predictive standard error (PSE) in (2.7b) is

$PSE_k = \hat{\sigma} N^{-1/2} H_k^{1/2} \leq \hat{\sigma} 0.183 \times 10$ for any $k$,

when $\mathcal{M}(1)$ or $\mathcal{M}(2)$ is used. In other words, all the individual PSEs are controlled below $1.83\hat{\sigma}$ whatever the standard error $\hat{\sigma}$ of the model maybe. This upper bound can be smaller by choosing smaller thresholds $c_q$ and $d_R$, but in such a case, the set of available variables to be selected in each model becomes smaller, which may lead to a larger $\hat{\sigma}$.

Though the XMSP does not use $y^o$, the adjusted CDs of regressing $y^o$ on $M(1)$ and $M(2)$ are respectively 0.77 and 0.73. On the other hand, the CD is 0.98 when $y^o$ is regressed on $M(1) \cup \{X4\}$, which implies a strong collinearity in the regressors because the CD is 0.98 when X4 is regressed on $M(1)$.

5.2 VR-Algorithm via PCA

Using the same data, let us consider the VR-Algorithm to get an *IC*- controlled class of models. Since the *I*-controlled class of variables has been obtained for $c_q = 0.90$ as the set in (5.1), the VR-Algorithm is here applied to the real data set in (1) and then to the sets of simulated data for (5.1) in (2) and (3). In the case of VI-Algorithm, $c_q$ and $d_R$ are the control parameters to be prespecified in advance, while in the case of VR-Algorithm it is necessary to prespecify $a$ in $G_{maz}$ and $b$ in (4.14) in addition to $c_q$ and $d_R$. Here, we consider the case of $a = 0.90$ and 0.95 and $b = 0.4$ together with $c_q = 0.9$ and $d_R = 0.9$.

(1) The real data case of X(1,0.9)

To treat the data in the case of $c_q = 0.9$ in (5.3), rename the variables in the set (5.1) by

(5.5)     X(1,0.9)={X2, X4, X7, X12}≡ $\{X(2), X(3), X(4), X(5)\}$.

Under this notation of the variables, a PCA is made to derive relevant statistics required in the VR-Algorithm. For example, $Z^* = (z_{(2)}, z_{(3)}, z_{(4)}, z_{(5)})$ denotes the $N \times 4$ matrix of standardized variables of $\{X(2), X(3), X(4), X(5)\}$.

Table 5.3 summarized analytical results on two cases of (a) real data and (b) simulated data via PCA to obtain a *C*-controlled class of models in X(1,0.9). The sub-tables of 1A, 1B and 1C in Table 5.3 respectively give

1A) the correlation matrix $Z^{*'}Z^*$ computed from the original data,
1B) the orthogonal matrix $P = [p_{(2)}, \cdots, p_{(5)}]$ where $p_{(j)}$ is the *j*-th latent vector corresponding to the *j*-th latent root $\lambda_j$ and
1C) the latent roots $\{\lambda_j\}$ of $Z^{*'}Z^*$ with $\lambda_2 > \lambda_3 > \lambda_4 > \lambda_5$ and relative contributions $\delta_j = \lambda_j/4$ $(j = 2, \cdots, 5)$.

The corresponding results are given in (2A,2B,2C) for a set of simulated data in Table 5.3. The simulation case will be discussed later.

Looking into the correlation matrix in Sub-table 1A, the correlation between X(2) and X(3) is 0.990, which is very high, though the two variables are not deleted in the *I*-controlling variable selection process. This will show the existence of "strong" collinearity among the 4 variables, implying that it will make a significant effect on the stability of some OLSEs of the regression model with the 4 variables plus X1. Sub-table 1B gives the matrix $\boldsymbol{P}$ and so $\boldsymbol{p}_{(2)} = (p_{(22)}, p_{(32)}, p_{(42)}, p_{(52)})'$ is the first column of the Sub-table 1B, and the largest latent root is $\lambda_2 = 3.188$ with $\delta_2 = 0.797$, implying that the 2nd PCC class is only selected for consideration into strong collinearity.

Here, using the notation $d(2,k) \equiv |\sqrt{\lambda_2} p_{k2}|$ of the 2nd PC collinearity identifier of the $k$-th variable in (4.12), the values are

(5.6a)  $d(2,2) = 0.980, d(2,3) = 0.977, d(2,4) = 0.733$ and $d(2,5) = 0.859$,

and so for $a = 0.90$, the set $G_{2az}$ in (4.13a) becomes

(5.6b)     $G_{2az} = \{\boldsymbol{z}_{(k)} \in \boldsymbol{Z}^* | d(2,k) \geq 0.90\} = \{\boldsymbol{z}_{(2)}, \boldsymbol{z}_{(3)}\}$.

Since $G_{jaz}$ is shown to be empty set $\emptyset$ for $a = 0.90$ and $j=3, 4, 5$, the two variables $\boldsymbol{z}_{(2)}$ and $\boldsymbol{z}_{(3)}$, which correspond to X2 and X4 by (5.5) respectively, cannot be used simultaneously according to the rule stated in Section 4.2. In fact, $\delta_2 = 0.797 > 0.4 > \delta_3 = 0.156$ and so only one PCC class is obtained. Therefore, the VR-Algorithm also selects the same two models $\{M(1), M(2)\}$ in (5.3) as model candidates for $IC$-controlled class of models. In fact, these models pass the screening procedure in 4.2 (4).

Table 5.3 1A shows that the correlation between $\boldsymbol{z}_{(2)}$ and $\boldsymbol{z}_{(3)}$ is 0.990, while the lower bound in (4.14b) for $|\boldsymbol{z}_2' \boldsymbol{z}_3|$ is $L_{2.2.3} = d(2,2) \times d(2,3) - (1 - a^2) = 0.826$ for $a = 0.95$. Here, note that since (5.6b) holds even for $a = \min\{d(2,2), d(2,3)\} = 0.977$ because of (5.6a), the lower bound for $|\boldsymbol{z}_2' \boldsymbol{z}_3|$ can be chosen as $L_{2.2.3} = 0.912$ for such $a$.

Table 5.3 VR-Algorithm. The sub-tables (1A,1B, 1C) and (2A,2B,2C) in real data and simulated data respectively gives correlation matrix, orthogonal matrix and latent roots in the original case and simulated case.

| 1A | X(2) | X(3) | X(4) | X(5) | 2A | X(2) | X(3) | X(4) | X(5) |
|---|---|---|---|---|---|---|---|---|---|
| X(2) | 1 | 0.990 | 0.640 | 0.824 | X(2) | 1 | 0.989 | 0.693 | 0.821 |
| X(3) | 0.990 | 1 | 0.653 | 0.801 | X(3) | 0.989 | 1 | 0.687 | 0.785 |
| X(4) | 0.640 | 0.653 | 1 | 0.395 | X(4) | 0.693 | 0.687 | 1 | 0.442 |
| X(5) | 0.824 | 0.801 | 0.395 | 1 | X(5) | 0.821 | 0.785 | 0.442 | 1 |
| 1B | $\boldsymbol{p}_{(2)}$ | $\boldsymbol{p}_{(3)}$ | $\boldsymbol{p}_{(4)}$ | $\boldsymbol{p}_{(5)}$ | 2B | $\boldsymbol{p}_{(2)}$ | $\boldsymbol{p}_{(3)}$ | $\boldsymbol{p}_{(4)}$ | $\boldsymbol{p}_{(5)}$ |
| X(2) | -0.549 | -0.100 | -0.406 | 0.724 | X(2) | -0.546 | 0.083 | -0.369 | 0.747 |
| X(3) | -0.547 | -0.062 | -0.473 | -0.688 | X(3) | -0.540 | 0.054 | -0.521 | -0.658 |
| X(4) | -0.411 | 0.839 | 0.357 | 0.005 | X(4) | -0.429 | -0.811 | 0.396 | -0.028 |
| X(5) | -0.481 | -0.531 | 0.696 | -0.048 | X(5) | -0.475 | 0.576 | 0.660 | -0.085 |
| 1C | 1 | 2 | 3 | 4 | 2C | 1 | 2 | 3 | 4 |
| $\lambda$ | 3.188 | 0.625 | 0.178 | 0.010 | $\lambda$ | 3.236 | 0.570 | 0.185 | 0.009 |
| $\delta$ | 0.797 | 0.156 | 0.045 | 0.002 | $\delta$ | 0.809 | 0.142 | 0.046 | 0.002 |

(2) The simulated data case with $c_q = 0.90$

Next, let us consider the case of simulation data. Since regression matrix $\boldsymbol{X}$ is of great variety in its composition of variables, it is not easy to set up a typical case for simulation.

In fact, $X$ itself is often fixed in regression model and collinearity in $X$ is often regarded as a matter of the given data. In practice, some variables apparently causing the strong collinearity are often modified or deleted. In VR-Algorithm, those variables are identified and used in separated models of different combinations of variables.

Another problem in simulation is that even if $X$ is generated by a distribution, it is too complex to specify it according to the observed data. Among the variables in $X$, there may be such one-time variables as structural changes or COVID-19 dummy variables. In the example of the automobiles' gasoline consumption, there are a dummy variable (X12), two ratio variables (X5, X6), a discrete variable (X7) and other continuous variables in $X$. Here, without taking these factors into account, we simply generate data with the assumption that

(5.7)   $w_n$ is *iid* as normal distribution $N_4(\mathbf{0}, \boldsymbol{\Phi})$,

where $w_n$ represents the *n*-th observation of $(X(2), X(3), X(4), X(5))$, and $N_4(\mathbf{0}, \boldsymbol{\Phi})$ denotes normal distribution with mean $\mathbf{0}$ and covariance matrix $\boldsymbol{\Phi}$, which is assumed to be the correlation matrix given in 1A of Table 5.3. Then by the simulated data $\{w_j: j = 1, \cdots, N\}(N = 50)$, we obtain a simulated matrix $Z_s^*$ of standardized variables.

Under this setting, $Z_s^*$ is similarly analyzed along the case of real data in Table 5.3. The results are summarized in the Sub-tables 2A, 2B and 2C. The correlation matrix in 2A that is computed with simulated data is similar to the original one in 1A, though the correlation between X(2) and X(4) is a bit different. Then, in the same way as the above case, it is easy to obtain almost same results

$$G_{2az} = \{z_{(k)} \in Z_s^* | d(2, k) \geq 0.90\} = \{z_{(2)}, z_{(3)}\},$$

and $G_{jaz}$ is shown to be empty set $\emptyset$ for $a = 0.90$ and $j=3, 4, 5$. Therefore, the VR-Algorithm also selects the same two models $\{M(1), M(2)\}$ in (5.3) even for this simulated data. This result will imply that, although the simulated data type is quite different, the decision threshold $a = 0.90$ for the second PCC class of variables well picks up strongly collinear variables $\{z_{(2)}, z_{(3)}\}$ in the class $G_{2az}$ under the assumption in (5.7). It is natural to have such a result under (5.7) since the 2nd PC identifier $d(2, k) = |\sqrt{\lambda_2} p_{(k2)}|$ for threshold is a function of $Z_s^{*'} Z_s^*$ as a decision statistic. In fact, it measures the correlation between the *k*-th variable $z_{(k)}$ and the 2nd normalized PC of $Z_s^{*'} Z_s^*$.

(3) The simulation analysis on the functions of the 2nd PCC class

Now, it may be interesting to ask what happens to the result of selected variables in the VR-Algorithm if the above simulation process is repeated under the assumption of normality in (5.7) with the same correlation matrix. To get an answer, we generated 1,000 of $Z_s^*$'s under (5.7) to look into which variables the 2nd PCC class $G_{2az}$ includes or "selects" relative to the collinearity threshold $a$ when $N = 50$. Here, the threshold is expected to control the strong collinearity among the variables. The following Table 5.4 lists the 4 groups {G1,G2,G3,G4} of variables that $G_{2az}$ selected in 1000 simulations and it gives the frequencies of the groups that $G_{2az}$ selected for the two cases: $a = 0.90$ and $a = 0.95$, where the original names of the variables in Table 5.1 are used. It is noted that G1= {X2,X4} is included in all the groups. Also, the variables in $G_{2az}$ are required to be not included more than one variable in each model if they are selected.

First, consider the case of $a = 0.90$. Then, in the group G4 of Table 5.4, $G_{2az}$ contains variables X2, X4, X7 and X12 as a combination of strong collinearity. If $G_{2az}$ selects G4 with probability 1/1000 under the condition that the data really follows (5.7), the four variables are required to be separately used in each model according to the VR-Algorithm. In this rare case there are the four model candidates: M(X1,Xk) (k=2,4,7,12), each of which is a simple regression model. Such a rare case may happen even in real world, and then it may be better to check the data for the causes. But if it really happened, we will have to use these simple regression models, because first of all, the observed data $X$ is of that structure in regression model, second we do not know the probability that it happened and mostly we cannot get an alternative data.

Also, in G3, there would be the three model candidates: M(X1,X2,X12), M(X1,X4,X12) and M(X1,X7,X12) with probability 1/1000, where X12 was originally dummy variable taking values of 0 or 1, though it is here assumed to be normal. While, in G2, there will be the three model candidates: M(X1,X2,X7), M(X1,X4,X7) and M(X1,X7,X12) with probability 104/1000. Thus, when $a = 0.90$, this case happens with probability 0.104 under the assumption of normality with the specific correlation matrix in Sub-table 1A. Even if it happens in real data, we have to use the models as given so long as the selection procedure has been accepted, though nobody knows how the data is generated in reality. The case of G1 occurs with probability 894/1000 and it is similar to the real data case.

Table 5.4 The frequency of the selected combinations of variables in $G_{2az}$

|  | G1 | G2 | G3 | G4 |
|---|---|---|---|---|
|  | X2,X4 | X2,X4,X12 | X2,X4,X7 | X2,X4,X7,X12 |
| $a = 0.90$ | 894 | 104 | 1 | 1 |
| $a = 0.95$ | 1000 | 0 | 0 | 0 |

However, if the threshold level $a$ is made lifted to $a = 0.95$ more stringently for the 2nd PCC class $G_{2az}$ of variables, the case of G1 in Table 5.4 shows the same result as the case of real data in 5.2 (1) with probability 1. These model candidates pass the screening process in 4.2 (4) to get a class of *IC*-controlled models.

This simulation result will show that it will be better to choose a higher threshold $a$ in order not to select variables that have mild collinearity with the PC variables but to select variables that have really strong collinearity with them.

5  Conclusion

By the concept of EEM-M, we developed a new integrated process of bundle model – XMSP - $y^o$MSP to obtain an empirical effective model for given data ($\boldsymbol{y^o}, \boldsymbol{X}$). The XMSP we proposed selects a class of models with inefficiency-controlled and collinearity-controlled OLS estimates in each model without using $\boldsymbol{y^o}$. The property of the estimates corresponds to inefficiency measure and collinearity measure associated with individual predictive sampling variances. Using these two measures of individual estimates in each estimated model, we defined the IC Risk Index, by which models are made to be compared in the XMSP. Finally, to materialize our conceptual and analytical results, we proposed two algorithms to control collinearity and demonstrated numerical analysis including simulations.

6  Acknowledgement

Kariya and Hayashi's portion of this work was supported by JSPS KAKENHI Grant Number 21K01431. Kurata's portion is partially supported by JSPS KAKENHI Grant Number 19K11853. The comments and suggestions of the reviewers were especially fruitful, and resulted in a thorough revision of this paper.

Appendix.   Proofs of Proposition 2.1 and Lemmas 4.1 and 4.2.

(1) Proof of Proposition 2.1. By the definition of $\check{R}_k^2$, let

$$\breve{M}_k = \breve{X}_k(\breve{X}_k'\breve{X}_k)^{-1}\breve{X}_k' \text{ with } \breve{X}_k = \{x_j \in X, j \neq k\},$$
$$\hat{x}_k = \breve{M}_k x_k \text{ and } M_e = e(e'e)^{-1}e'.$$

Then, using $\breve{M}_k e = e$,

(A.1)     $\check{R}_k^2 = s_{\hat{x}k}^2/s_{xk}^2 = \hat{x}_k'(I - M_e)\hat{x}_k / x_k'(I - M_e)x_k.$

Hence, $\check{R}_k^2 = 0$ if and only if $x_k'(\breve{M}_k - M_e)x_k$=0. Therefore, for any *k,* $H_k = 1$ if and only if $\bar{x}_k = 0$ and $x_k'\breve{M}_k x_k = x_k'M_e x_k$ =0  since $N\bar{x}_k = e'x_k$, which in turn holds if and only if $\bar{x}_k = 0$ and $\breve{X}_k'x_k = \mathbf{0}$, implying $x_k'x_j = 0$  $(k \neq j)$. Thus, the result follows.

(2) Proof of Lemma 4.1.

As is well known, $\check{R}_k^2$  is obtained as its CD via the regression model

$$x_k = \sum_{j=1, j\neq k}^{K} \theta_j x_j + v_k.$$

Then as is well known, the CD of this model is equal to that of the model:

(A.2)     $x_k = \theta_1^* e + \sum_{j=2, j\neq k}^{K} \theta_j \sqrt{N} s_{xj} z_j + v_k$ with

(A.3) $\quad \mathbf{z}_j = (z_{nj})$ and $z_{nj} = (x_{nj} - \bar{x}_j)/\sqrt{N}s_{xj}$ $(j \geq 2)$,

where $\mathbf{x}_1 = \mathbf{e}$ and $\theta_1^* = (\theta_1 + \sum_{j \neq k} \theta_j \bar{x}_j)$. Hence, as far as the CD is concerned, we can assume $\bar{x}_j = 0$ for any $j \geq 2$ $(j \neq k)$ without loss of generality. Assuming it, the observations to compute $\check{R}_k^2(X)$ are expressed as $\mathbf{x}_j = \mathbf{z}_j \sqrt{N} s_{xj} (j \neq k)$, and $\mathbf{x}_k = \bar{x}_k \mathbf{e} + \mathbf{z}_k \sqrt{N} s_{xk}$, and so the whole $\mathbf{X}$ is expressed as

(A.4) $\quad \mathbf{X} = [\mathbf{e}, \mathbf{X}^*] = [\mathbf{e}, \mathbf{O}_k] + [\mathbf{0}, \mathbf{Z}^*] diag\{0, \mathbf{D}\}$,

where $\mathbf{O}_k$ is $N \times (K-1)$ zero matrix except for the $k$-th column $\bar{x}_k \mathbf{e}$, $\mathbf{Z}^* = (\mathbf{z}_2, \cdots, \mathbf{z}_K): N \times (K-1)$ with $\mathbf{z}_k' \mathbf{e} = 0$ and $\mathbf{D} = diag\{\sqrt{N}s_{x2}, \cdots, \sqrt{N}s_{xK}\}$. Let $\check{R}_k^2(\mathbf{Z}^*)$ be the CD of $\mathbf{z}_k$ when $\mathbf{z}_k$ is regressed on $\check{\mathbf{Z}}_k^*$, which is $\mathbf{Z}^*$ with the $k$-th variable deleted. Note $\mathbf{e}'\mathbf{Z}^* = \mathbf{0}$, implying that the mean of each $\mathbf{z}_j$ is zero.
Based on (A.4), we prove $\check{R}_k^2(\mathbf{X}) = \check{R}_k^2(\mathbf{Z}^*)$ $(k \geq 2)$.

To express $\check{R}_k^2(\mathbf{X})$ in terms of $\mathbf{Z}^*$, let $\tilde{M}_{Xk} = \check{\mathbf{X}}_k (\check{\mathbf{X}}_k' \check{\mathbf{X}}_k)^{-1} \check{\mathbf{X}}_k'$ with $\check{\mathbf{X}}_k = \{\mathbf{x}_j \in \mathbf{X}, j \neq k\}$, $\mathbf{x}_j' \mathbf{e} = 0$ as $\bar{x}_j = 0$, $\hat{\mathbf{x}}_k = \tilde{M}_{Xk} \mathbf{x}_k$ and $M_e = \mathbf{e}(\mathbf{e}'\mathbf{e})^{-1}\mathbf{e}'$ with $\tilde{M}_{Xk}\mathbf{e} = \mathbf{e}$. Then

(A.5) $\quad \check{R}_k^2 = s_{\hat{x}k}^2 / s_{xk}^2 = \frac{\hat{\mathbf{x}}_k'(I-M_e)\hat{\mathbf{x}}_k}{\mathbf{x}_k'(I-M_e)\mathbf{x}_k} = \frac{\mathbf{x}_k'(\tilde{M}_{Xk}-M_e)\mathbf{x}_k}{\mathbf{x}_k'(I-M_e)\mathbf{x}_k}$.

Since $\mathbf{x}_k = \bar{x}_k \mathbf{e} + \mathbf{z}_k \sqrt{N} s_{xk}$ and $\mathbf{z}_k'\mathbf{z}_k = 1$, $\check{R}_k^2 = \mathbf{z}_k' \tilde{M}_{Xk} \mathbf{z}_k$, where $(\tilde{M}_{Xk} - M_e)\bar{x}_k \mathbf{e} = \mathbf{0}$ is used. Here, without loss of generality, let $k=K$ for notational convenience. Also write $\check{\mathbf{X}}_K = [\mathbf{e}, \mathbf{x}_2, \mathbf{x}_3, \cdots, \mathbf{x}_{K-1}] = [\mathbf{e}, \mathbf{O}] + [\mathbf{0}, \check{\mathbf{Z}}_K^*] diag\{0, \check{\mathbf{D}}_K\}$, $\check{\mathbf{Z}}_K^* = [\mathbf{z}_2, \cdots, \mathbf{z}_{K-1}]$ and $\check{\mathbf{D}}_K = diag\{\sqrt{N}s_{x2}, \cdots, \sqrt{N}s_{xK-1}\}$. Here $\mathbf{O}$ is the zero matrix. Then,

(A.6) $\quad \mathbf{z}_K' \tilde{M}_{XK} \mathbf{z}_K = (0, \mathbf{z}_K' \check{\mathbf{Z}}_K^* \check{\mathbf{D}}_K) \begin{bmatrix} N & \mathbf{O}' \\ \mathbf{O} & \check{\mathbf{D}}_K \check{\mathbf{Z}}_K^{*'} \check{\mathbf{Z}}_K^* \check{\mathbf{D}}_K \end{bmatrix}^{-1} (0, \mathbf{z}_K' \check{\mathbf{Z}}_K^* \check{\mathbf{D}}_K)'$

$\quad\quad\quad\quad = \mathbf{z}_K' \tilde{M}_{ZK} \mathbf{z}_K$.

This completes the proof.

(3) Proof of Lemma 4.2.

Let $\varphi_{2kj} = \lambda_3 p_{3k} p_{3j} + \cdots + \lambda_K p_{Kk} p_{Kj}$. Then from (4.11b), $\varphi_{2kk} = 1 - \lambda_2 p_{2k}^2 \leq 1 - a^2$ and from (4.11a), $\mathbf{z}_k' \mathbf{z}_j - \lambda_2 p_{2k} p_{2j} = \varphi_{2kj}$. Hence, applying the Cauchy-Schwarz Inequality to $\varphi_{2kj}$, we obtain

$\quad |\mathbf{z}_k' \mathbf{z}_j - \lambda_2 p_{2k} p_{2j}| \leq [\varphi_{2kk} \cdot \varphi_{2jj}]^{1/2} \leq 1 - a^2$.

This implies that $\pm \lambda_2 p_{2k} p_{2j} - (1 - a^2) \leq \pm \mathbf{z}_k' \mathbf{z}_j \leq \pm \lambda_2 p_{2k} p_{2j} + (1 - a^2)$ where the $+$ sign or $-$ sign in the double sign $\pm$ should be selected respectively according to the sign of $p_{2k} p_{2j} \geq 0$ or $p_{2k} p_{2j} < 0$. Also, (4.14b) follows as $\lambda_2 |p_{2k} p_{2j}| \geq a^2$.